\documentclass[a4paper,11pt]{article}
\usepackage{amsmath,amssymb,color,comment}
\usepackage{slashed, tensor, bm, physics}
\usepackage{caption}
 \usepackage{graphicx}
\usepackage{float}
\usepackage[compat=1.1.0]{tikz-feynhand}
\usepackage{jheppub} % for details on the use of the package, please
\usepackage[T1]{fontenc} % if needed
\usepackage{booktabs} % to make tables look cool. (added by TA)

\title{Composite dark matter with forbidden annihilation}
\author[a]{Tomohiro Abe,}
\author[b]{Ryosuke Sato,}
\author[b]{and Takumu Yamanaka}

\affiliation[a\,]{Department of Physics and Astronomy, Faculty of Science and Technology, \\Tokyo University of Science, Yamazaki, Noda, Chiba 278-8510, Japan}
\affiliation[b\,]{Department of Physics, Osaka University, \\Machikaneyama-cho, Toyonaka, Osaka 560-0043, Japan}

\emailAdd{abe.tomohiro@rs.tus.ac.jp}
\emailAdd{rsato@het.phys.sci.osaka-u.ac.jp}
\emailAdd{yamanaka@het.phys.sci.osaka-u.ac.jp}

\abstract{
A dark matter model based on QCD-like $SU(N_c)$ gauge theory with electroweakly interacting dark quarks is discussed. Assuming the dark quark mass $m$ is smaller than the dynamical scale $\Lambda_d \sim 4\pi f_d$, the main component of the dark matter is the lightest $G$-parity odd dark pion associated with chiral symmetry breaking in the dark sector. We show that nonzero dark quark mass induces the universal mass contribution to both $G$-parity odd and even pions, and their masses tend to be degenerate. As a result, dark pion annihilation into heavier $G$-parity even dark pion also affects the dark matter relic abundance. Thus, our setup naturally accommodates forbidden dark matter scenario and realizes heavy dark matter whose mass is ${\cal O}(1$--$100)~{\rm TeV}$, which is different from conventional electroweakly interacting dark matter such as minimal dark matter. We also discuss CP-violation from the $\theta$-term in the dark gauge sector and find that the predicted size of the electron electric dipole moment can be as large as $\sim 10^{-32}~e~{\rm cm}$.
}

\begin{document} 
\begin{flushright}
OU-HET-1219
\end{flushright}
\maketitle
\flushbottom
\section{Introduction}
There is much evidence that almost a quarter of the energy in our current universe consists of dark matter (DM) \cite{Planck:2018vyg}. A new stable particle in an extension of the Standard Model (SM) can be a candidate for DM, and the thermal freeze-out scenario is one of the most attractive scenarios for the production of particle DM in the early universe. In this scenario, the DM interacts with the SM particles and couples with thermal plasma in the early universe. Direct detection experiments can probe this interaction. There have been extensive efforts for this search, and particularly the DM whose mass is ${\cal O}(1$--$100)~{\rm GeV}$ is severely constrained by direct detection experiments such as LZ \cite{LZ:2022lsv} and XENONnT \cite{XENON:2023cxc}.

A possible direction under this situation is to explore heavy DM whose mass is ${\cal O}(1$--$100)$ TeV, which is less constrained by the direct detection experiments.
The upper bound on the DM mass comes from the unitarity bound of its annihilation cross section \cite{Griest:1989wd}.
Recently, there has been significant progress on indirect detection experiments such as HAWC \cite{HAWC:2019jvm}, HESS \cite{HESS:2018cbt}, MAGIC \cite{MAGIC:2022acl}, and VERITAS \cite{Acharyya:2023ptu}, and those experiments started to be sensitive to the annihilation of DM whose mass is above ${\cal O}(1)$ TeV. Furthermore, near-future observations, CTA \cite{CTA:2020qlo}, LHAASO \cite{LHAASO:2019qtb}, and SWGO \cite{Abreu:2019ahw} are expected to improve sensitivity on the DM annihilation \cite{Rinchiuso:2020skh} (for a review, see, e.g., \cite{Boddy:2022knd}.) Thus, we have a chance to find such a heavy DM.
In the thermal freeze-out scenario, heavier DM requires a larger coupling to be consistent with the measured DM relic abundance. One natural way to accommodate the large coupling is to assume that DM is a composite particle from new strong dynamics.

In this paper, we discuss heavy composite particles as a DM candidate. For a review, see refs.~\cite{Kribs:2016cew, Cacciapaglia:2020kgq, Cline:2021itd}. In particular, we focus on a dark pion DM arising from chiral symmetry breaking in a new QCD-like sector \cite{Hur:2007uz,Bai:2010qg, Hur:2011sv, Buckley:2012ky, Bhattacharya:2013kma, Hietanen:2013fya, Bellazzini:2014yua, Hochberg:2014dra, Antipin:2014qva, Antipin:2015xia, Hatanaka:2016rek, Mitridate:2017oky,Choi:2018iit, Redi:2018muu, Dondi:2019olm, Contino:2020god, Alexander:2023wgk}.
We assume dark quarks are charged under the electroweak gauge symmetry, and the dark and SM sectors can communicate via electroweak gauge interaction.
Dark quark condensation provides the chiral symmetry breaking and (pseudo) Nambu-Goldstone (NG) bosons, i.e., dark pions. The lightest dark pion is stable because of $G$-parity and is a good candidate for DM, as pointed out in ref.~\cite{Bai:2010qg}. We found that the mass term of the dark quarks significantly affects the mass spectrum of the dark pions and naturally accommodates the forbidden dark matter scenario \cite{Griest:1990kh, DAgnolo:2015ujb}. Also, the $\theta$-term in the dark $SU(N_c)$ gauge sector induces CP-violating interaction and the electron electric dipole moment (EDM).

This paper is organized as follows. In section \ref{sec:models}, we construct a chiral Lagrangian of the dark pions and discuss their mass spectrum. We discuss the DM relic abundance and the electron EDM in section \ref{sec:phenomenology}.
In section~\ref{sec:signals}, we briefly discuss phenomenological aspects of direct detection experiments,  indirect detection experiments, collider experiments, and gravitational wave (GW) observations.
Section~\ref{sec:conclusion} is dedicated to presenting our conclusions.
In appendix~\ref{sec:wzw}, we briefly show derivation of the Wess-Zumino-Witten term.

\section{Model}\label{sec:models}
\noindent
In this section, we describe a DM model based on a new QCD-like gauge theory. Our model is an explicit example of models that were originally discussed in ref.~\cite{Bai:2010qg}.
We introduce an $SU(N_c)$ gauge symmetry ($N_c \geq 3$) and new Weyl fermions, $\psi$ and $\bar{\psi}$, as the dark quarks and dark anti-quarks which are charged under the $SU(N_c)$ and the SM gauge group, $SU(3)_c\times SU(2)_W\times U(1)_Y$. The charge assignments are shown in table~\ref{tab:darkquark}.
For a concrete discussion, in this paper, we introduce three flavors of $\psi$ and $\bar\psi$, and they form triplets under the $SU(2)_W$ gauge symmetry. We assume they are not charged under the $SU(3)_c$ and $U(1)_Y$ gauge symmetry. 
The renormalizable Lagrangian is given by 
\begin{align}
  \mathcal{L} = \mathcal{L}_{\text{SM}} &+ \bar{\Psi}_i(i\slashed{D}-m)\Psi_i-\frac{1}{4}G_{\mu\nu}^AG^{A \mu\nu} + \frac{g_d^2\theta}{32\pi^2} G_{\mu\nu}^A \tilde G^{A\mu\nu} , \quad 
  \Psi_i \equiv \mqty(\psi_i\\ \bar{\psi}^{\dagger}_i),
  \label{eq:renormlag}
\end{align}
where $i=1,2,3$ is the flavor index, $m$ is the mass of the dark quarks, the third term is the kinetic term of $SU(N_c)$ gauge field, $g_d$ is the gauge coupling of $SU(N_c)$, and $\theta$ is the vacuum phase of $SU(N_c)$ gauge sector. We take $-\pi < \theta \leq \pi$. Note that $\theta$ has a physical effect if $m$ is nonzero.
The $SU(N_c)$ gauge interaction becomes strong at low energy and confinement takes place at the dynamical scale $\Lambda_d$. In this paper, we are interested in the case of the dark quark mass $m \lesssim \Lambda_d$.
The massless case has been discussed by the previous literature including \cite{Bai:2010qg}, and the other limit $m \gg \Lambda_d$ has been discussed in ref.~\cite{Mitridate:2017oky}.
For later purposes, we absorb the CP-phase into the phase of the mass term by chiral phase rotation of the $\Psi_i$ field: 
\begin{align}
\Psi_i \to \exp\left( -\frac{i\theta}{6} \gamma^5 \right) \Psi_i.
\end{align}
Then, we obtain
\begin{align}
  \mathcal{L} = \mathcal{L}_{\text{SM}} &+ \bar{\Psi}_i \left[ i\slashed{D}-m \exp\left( -\frac{i\theta}{3}\gamma^5 \right) \right] \Psi_i-\frac{1}{4}G_{\mu\nu}^AG^{A \mu\nu}.
  \label{eq:renormlag2}
\end{align}
\begin{table}[tbp]
  \centering
  %\begin{tabular}{|c||c|c|c|c|}
  \begin{tabular}{ccccc}
    \toprule %\hline
     Field & $SU(N_c)$ & $SU(3)_c$ & $SU(2)_W$ & $U(1)_Y$\\
    \midrule  %\hline\hline
      $\psi$ & $N_c$ & $\bm{1}$ & $\bm{3}$ & 0\\
      \hline
      $\bar{\psi}$ & $\bar{N_c}$ & $\bm{1}$ & $\bm{3}$ & 0\\
     \bottomrule % \hline
 \end{tabular}
  \caption{The charge assignments of the dark quarks under $SU(N_c)$ and the SM gauge group, $SU(3)_c\times SU(2)_W\times U(1)_Y$.}
  \label{tab:darkquark}
\end{table}
\subsection{Chiral Lagrangian}
Let us discuss the chiral symmetry of the Lagrangian.
The current Lagrangian has an approximate $SU(3)_L\times SU(3)_R$ global symmetry.
To be more specific, $\psi_i$ and $\bar{\psi}^\dagger_i$ transform under this global symmetry as
\begin{align}
 \psi_i \to L_{ij} \psi_j, \quad
 \bar{\psi}_i^\dagger \to R_{ij} \bar{\psi}_j^\dagger,
\end{align}
where $L$ and $R$ are $3\times 3$ unitary matrices of the $SU(3)_L$ and $SU(3)_R$ global transformations, respectively. The $SU(2)_W$ gauge symmetry is embedded into the diagonal subgroup $SU(3)_V \subset SU(3)_L \times SU(3)_R$, and the field transforms as 
\begin{align}
 \Psi_j \to [\exp(i T^a \theta^a)]_{jk} \Psi_k,
\end{align}
where
\begin{align}
  T^a \theta^a = \mqty( 
  0 & -i\theta^3 & i\theta^2\\
  i\theta^3 & 0 & -i\theta^1\\
  -i\theta^2 & i\theta^1 & 0 ).
\end{align}
The $SU(3)_L \times SU(3)_R$ symmetry is explicitly broken by the $SU(2)_W$ gauge coupling and the dark quark mass $m$.

The $SU(N_c)$ gauge interaction is asymptotically free, and confinement occurs at the dynamical scale $\Lambda_d$. As in the case of the real QCD, the dark quarks condense 
\begin{align}
  \langle {\bar\psi} \psi \rangle = -v_d^3\,\bm{1}.\label{eq:condensatevacuum}
\end{align}
Then the approximate $SU(3)_L\times SU(3)_R$ global symmetry is spontaneously broken into the diagonal subgroup $SU(3)_V$, in which the $SU(2)_W$ gauge group is embedded.
The low energy effective theory below the dynamical scale $\Lambda_d$ is described by $3\times 3$ special unitary matrix field $U$, which is transformed as $U \to L U R^\dagger$ under the $SU(3)_L\times SU(3)_R$ transformation.
The pion fields associated with the chiral symmetry breaking are embedded into $U$.
The dark quark mass matrix $M \equiv \text{diag}(m,m,m)e^{i\theta/3}$ can be regarded as a spurion field which is transformed as $M \to RML^\dagger$.
Then we can construct ${\cal O}(p^2)$ chiral Lagrangian as 
\begin{align}
  \mathcal{L}_2 &= \frac{f_d^2}{4}\Tr[D_{\mu}UD^{\mu}U^{\dagger}] - V(U), \label{eq:chirallag}\\
  V(U) &\equiv -v_d^3\Tr[MU+\mathrm{h.c.}], \label{eq:meson potential}
\end{align}
where $D_{\mu}U =\partial_{\mu}U-ig[W_{\mu}, U]$, 
$g$ is the $SU(2)_L$ gauge coupling, 
and 
\begin{align}
  W_{\mu} = \mqty( 0 & -iW_{\mu}^3 & iW_{\mu}^2\\
  iW_{\mu}^3 & 0 & -iW_{\mu}^1\\
  -iW_{\mu}^2 & iW_{\mu}^1 & 0 ). \label{eq:Wmu matrix}
\end{align}
In the following of this paper, we utilize the chiral Lagrangian eq.~(\ref{eq:chirallag}) for the analysis of the vacuum and the dynamics of the dark pions. The chiral Lagrangian is a powerful effective theory to analyze low energy physics. The only assumption is the symmetry breaking pattern and the chiral symmetry breaking with small quark mass is well understood. See, e.g., ref.~\cite{Faber:2017alm} and reference therein for lattice QCD analysis and refs.~\cite{Coleman:1980mx, Veneziano:1980xs} for large $N_c$ analysis. We treat the dark quark mass as a perturbative parameter, i.e., the dark quark mass should be small enough compared to the dynamical scale. This shows the limitation of our analysis and it will be discussed in section \ref{sec:results}.
Note that $v_d$ and $f_d$ are related as $v_d^3\sim 4\pi f_d^3/\sqrt{N_c}$ by naive dimensional analysis (NDA) for counting the factor of $4\pi$ and $N_c$ \cite{Manohar:1983md, Georgi:1986kr, Luty:1997fk, Cohen:1997rt, Hill:2002ap}.

\subsubsection{The global minimum of the potential $V(U)$}
Let us investigate the global minimum of the potential $V(U)$ given in eq.~(\ref{eq:meson potential}).
The analysis on a similar potential has been presented, for example, in ref.~\cite{Smilga:1998dh}.
We denote a classical and uniform configuration of the field $U$ as $\expval{U}$.
In this subsection, we show that the potential energy is minimized at $\expval{U} = \bm{1}$ for any values of $m$ and $\theta \in (-\pi,\pi]$.

\begin{figure}[tbp]
  \centering
  \includegraphics[width=0.8\hsize]{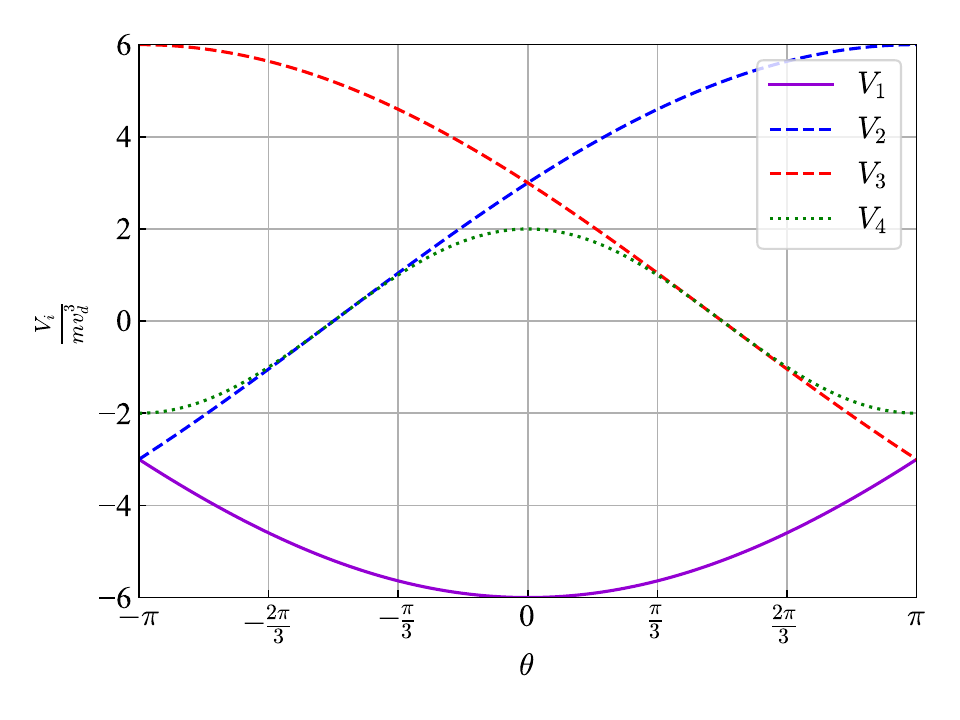}
  \caption{$V_i$'s given in eq.~(\ref{eq:V1V2V3V4}) as a function of $\theta$.}
  \label{fig:potential}
\end{figure}

Since $V(U)$ depends only on ${\rm tr}U$ and its complex conjugate, it is enough to investigate cases with diagonal $\langle U \rangle$. Note also that $U^\dagger U = 1$ and $\det U =1$. Then, we take
\begin{align}
   \langle U\rangle &\equiv \mqty(\dmat{e^{i\phi_1},e^{i\phi_2},e^{i\phi_3}}),\label{eq:pionphase}\\
  \phi_1 +\phi_2&+\phi_3=2\pi n, \quad n\in \mathbb{Z}.\label{eq:phasecnstr}
\end{align}
To find the global minimum of the potential, let us find stationary points in $V(U)$ under the constraint (\ref{eq:phasecnstr}). For this purpose, we introduce a Lagrange multiplier $\lambda$ and define $\tilde{V}$ as
\begin{align}
  \tilde{V} &\equiv V(\langle U\rangle)+\lambda(\phi_1+\phi_2+\phi_3-2\pi n),\\
  V(\langle U\rangle) &= -2mv_d^3 \left[ \cos\left(\frac{\theta}{3} + \phi_1\right) + \cos\left(\frac{\theta}{3} + \phi_2\right) + \cos\left(\frac{\theta}{3} + \phi_3\right) \right].
\end{align}
The stationary conditions on $\tilde V$ are given as
\begin{align}
  \pdv{\tilde{V}}{\phi_i} = 2mv_d^3\sin\left( \frac{\theta}{3} + \phi_i \right)+\lambda=0 \quad (i=1,2,3).
\end{align}
There are six independent solutions of them:
\begin{align}
  (\phi_1, \phi_2, \phi_3) =&\left(\frac{2\pi}{3}\ell, \frac{2\pi}{3}\ell, \frac{2\pi}{3}\ell\right)\quad (\ell=0,1,2),\label{eq:minsol1}\\
  & \left( \frac{2}{3}\theta, \frac{2}{3}\theta, -\frac{4}{3}\theta \right),\quad
   \left( \frac{2}{3}\theta, -\frac{4}{3}\theta, \frac{2}{3}\theta \right),\quad
    \left( -\frac{4}{3}\theta, \frac{2}{3}\theta, \frac{2}{3}\theta \right).
  \label{eq:minsol2}
\end{align}
Equivalently, $\langle U \rangle$ at the stationary points are given as
\begin{align}
  \langle U \rangle = &\text{diag}(e^{i\frac{2\pi}{3}\ell}, e^{i\frac{2\pi}{3}\ell} , e^{i\frac{2\pi}{3}\ell}), \,\text{diag}(e^{2i\theta/3}, e^{2i\theta/3}, e^{-4i\theta/3}), \,\notag\\
&\text{diag}(e^{2i\theta/3}, e^{-4i\theta/3}, e^{2i\theta/3}), \,
\text{diag}(e^{-4i\theta/3}, e^{2i\theta/3}, e^{2i\theta/3}).
\end{align}
The three solutions in eq.~\eqref{eq:minsol2} give the same potential energy because the potential is symmetric under exchanges of $\phi_j$'s.
From these solutions, we obtain the following potential energies at the stationary points,
\begin{align}
  \begin{cases}
    V_1 = -6mv_d^3\cos\frac{\theta}{3} & (\text{from eq.~}\eqref{eq:minsol1}\text{ with } \ell=0),\\
    V_2 = -6mv_d^3\cos(\frac{\theta}{3}+\frac{2}{3}\pi) & (\text{from eq.~}\eqref{eq:minsol1}\text{ with } \ell=1),\\
    V_3 = -6mv_d^3\cos(\frac{\theta}{3}+\frac{4}{3}\pi) & (\text{from eq.~}\eqref{eq:minsol1}\text{ with } \ell=2),\\
    V_4 = 2mv_d^3\cos\theta. & (\text{from eq.~}\eqref{eq:minsol2}).
  \end{cases} \label{eq:V1V2V3V4}
\end{align}
The smallest value among these energies is the minimum value of $V(U)$.
The values of each $V_i$ are shown in figure~\ref{fig:potential}. 
We can see that $V_1$ is the smallest of $V_i$'s and $V(U)$ is minimized at $\langle U\rangle = \bm{1}$.
Note that the degeneracy of two vacua at $\theta = \pm\pi$ can be understood as the spontaneous breaking of CP symmetry known as Dashen's phenomenon \cite{Dashen:1970et, Gaiotto:2017yup}.
We can see that physical results are the same under $\theta \to \theta + 2\pi$ by choosing an appropriate vacuum though the chiral Lagrangian eq.~(\ref{eq:chirallag}) includes a factor of $\exp(i\theta/3)$.

\subsubsection{The effective Lagrangian for dark pions}
The fluctuation of $U$ around $U = \bm{1}$ can be interpreted as the dark pions.
Associated with the spontaneous symmetry breaking $SU(3)_L \times SU(3)_R \to SU(3)_V$, eight dark pions appear.
The dark pions form an octet of $SU(3)_V$ and they behave as triplet and quintuplet of $SU(2)_W \subset SU(3)_V$.
Throughout our discussion, we denote the triplet and quintuplet of the dark pions as $\chi$ and $\pi$, respectively.
$U$ can be decomposed into $SU(2)_W$ triplet $\Pi_3$ and quintuplet $\Pi_5$ as
\begin{align}
  U \equiv \exp(\frac{\sqrt{2}i}{f_d}(\Pi_3+\Pi_5)),
\end{align}
where $f_d$ is the decay constant of the dark pions.
$\Pi_3$ and $\Pi_5$ can be written as
\begin{align}
  \Pi_3 &\equiv \frac{1}{\sqrt{2}}\mqty(
    0 & -i\chi^0 & \frac{\chi^--\chi^+}{\sqrt{2}}\\
    i\chi^0 & 0 & -i\frac{\chi^-+\chi^+}{\sqrt{2}}\\
    -\frac{\chi^--\chi^+}{\sqrt{2}} & i\frac{\chi^-+\chi^+}{\sqrt{2}} & 0
  ),\label{eq:tripletmatrix}\\
  \Pi_5 &\equiv \mqty(
    \frac{\pi^0}{\sqrt{6}}-\frac{\pi^{++}+\pi^{--}}{2} & \frac{1}{2i}(\pi^{++}-\pi^{--}) & \frac{1}{2}(\pi^++\pi^{-})\\
    \frac{1}{2i}(\pi^{++}-\pi^{--}) & \frac{\pi^0}{\sqrt{6}}+\frac{\pi^{++}+\pi^{--}}{2} & -\frac{1}{2i}(\pi^+-\pi^{-})\\
    \frac{1}{2}(\pi^+ + \pi^{-}) & -\frac{1}{2i}(\pi^+ - \pi^{-}) & -\sqrt{\frac{2}{3}}\pi^0
  ).\label{eq:quintupletmatrix}
\end{align}
Here $\chi^0$ and $\pi^0$ are real scalar fields, and the negatively charged fields are defined as charge conjugation of the positively charged fields, such as $\chi^- = (\chi^+)^\dagger$, $\pi^- = (\pi^+)^\dagger$, and $\pi^{--} = (\pi^{++})^\dagger$.

In addition to these terms, the Wess-Zumino-Witten (WZW) terms arise from the chiral anomaly:
\begin{align}
  \mathcal{L}_{\text{WZW}}&\supset-\frac{g^2N_c}{16\sqrt{2}\pi^2f_{d}} \epsilon^{\mu\nu\rho\sigma}\Tr[\Pi_5W_{\mu\nu}W_{\rho\sigma}]\notag\\
  &\quad +\frac{i\sqrt{2}gN_c}{12\pi^2f_d^3}\epsilon^{\mu\nu\rho\sigma}\Tr[\partial_{\mu}\Pi_3\partial_{\nu}\Pi_3\partial_{\rho}\Pi_5W_{\sigma}+\partial_{\mu}\Pi_5\partial_{\nu}\Pi_5\partial_{\rho}\Pi_5W_{\sigma}]. \label{eq:wzw}
\end{align}
Here, we only pick up the leading terms of the pion fields $\Pi_i$'s. For a complete set of WZW terms, see Appendix \ref{sec:wzw}.
Hereafter, we consider the following chiral Lagrangian, 
\begin{align}
\mathcal{L}=\mathcal{L}_2+\mathcal{L}_{\text{WZW}}. \label{eq:Leff}
\end{align}

\subsection{Accidental symmetries and DM candidates}
The Lagrangian given in eq.~\eqref{eq:renormlag2} %eq.~(\ref{eq:Leff})
 has two unbroken accidental global symmetries which are relevant to DM candidates. One is $U(1)_\Psi$ symmetry. The Lagrangian is invariant under the following phase rotation:
\begin{align}
\Psi_i \to e^{i\alpha} \Psi_i.
\end{align}
This symmetry guarantees the stability of the lightest dark baryon $B$, which is a composite state of $N_c$ dark quarks. Thus, $B$ can be a candidate for DM.
The other accidental symmetry is $G$-parity symmetry:
\begin{align}
  \Psi_i\rightarrow \exp(i\pi T_2)\Psi_i^c,
\end{align}	
where $\Psi_i^c$ are charge conjugation of $\Psi_i$, and $T_2$ is a generator of $SU(2)_W$ transformation. This symmetry was originally discussed by Lee and Yang \cite{Lee:1956sw}, and later, it was applied to DM model building \cite{Bai:2010qg}. The lightest $G$-parity odd particle can also be a candidate for DM.

Let us discuss $G$-parity in our model in detail and identify the lightest $G$-parity odd particle. First, we identify charge conjugation for $SU(2)_W$ gauge field as
\begin{align}
W_\mu^1 \to -W_\mu^1, \quad
W_\mu^2 \to W_\mu^2, \quad
W_\mu^3 \to -W_\mu^3.
\end{align}
Equivalently, the charge conjugations of $W_{\mu}$ in the matrix form eq.~(\ref{eq:Wmu matrix}) is given as
\begin{align}
  W_{\mu}\rightarrow \mqty(\dmat{-1,1,-1})W_{\mu}\mqty(\dmat{-1,1,-1}).
\end{align}
To be consistent with this transformation law, $\psi$ and $\bar\psi$ should be transformed under the charge conjugation as
\begin{align}
  \mqty(\psi_1\\\psi_2\\\psi_3)\rightarrow \mqty(\bar{\psi}_1\\-\bar{\psi}_2\\\bar{\psi}_3),\quad \mqty(\bar{\psi}_1\\\bar{\psi}_2\\\bar{\psi_3})\rightarrow \mqty(\psi_1\\-\psi_2\\\psi_3).
\end{align}
Note that 
the operator $\exp(i\pi T_2)$ in the adjoint representation of $SU(2)_L$ is given by
\begin{align}
  \exp(i\pi T_2) = \mqty(\dmat{-1,1,-1}).
\end{align}
Then $W_\mu$ and the dark quarks $\psi,~\bar\psi$ behave under the $G$-parity transformation as
\begin{align}
W_\mu \to W_\mu, \quad
  \mqty(\psi_1\\\psi_2\\\psi_3)\rightarrow \mqty(-\bar{\psi}_1\\-\bar{\psi}_2\\-\bar{\psi}_3),\quad \mqty(\bar{\psi}_1\\\bar{\psi}_2\\\bar{\psi_3})\rightarrow \mqty(-\psi_1\\-\psi_2\\-\psi_3). \label{eq:gparity elementary}
\end{align}
The meson field $U_{ij}$ has the same charge as quark bilinear $\psi_i\bar\psi_j$.
Thus, under this $G$-parity transformation $U$ transforms as
\begin{align}
U \to U^T. \label{eq:gparity}
\end{align}
Then we can immediately see $\Pi_3$ and $\Pi_5$ transform as
\begin{align}
  \Pi_3 \to -\Pi_3, \qquad \Pi_5 \to \Pi_5.
\end{align}
Since $\Pi_3$ ($\Pi_5$) is anti-symmetric (symmetric), $\chi$ ($\pi$) is odd (even) under the $G$-parity transformation eq.~(\ref{eq:gparity}).
We can also see that the WZW term eq.~(\ref{eq:wzw}) also respects $G$-parity.
Because of $G$-parity, the lightest component of the triplet $\chi$ is stable and cannot decay. On the other hand, the quintuplet $\pi$ can decay into the electroweak gauge bosons via the WZW term, and its decay channels  into $\chi$ are kinematically forbidden.
Note that we can show that the invariance of ${\cal L}_{\rm WZW}$ under the $G$-parity transformation without expanding a series of pion field. For details, see appendix \ref{sec:wzw}. 
To summarize, in addition to the dark baryon, the lightest component of the triplet dark pion $\chi$ is stable and a DM candidate.

\subsection{Mass spectrum of dark pions}\label{sec:mass spectrum}
Let us estimate the mass spectrum of the dark pions.
There are three sources of the mass of the pions: 
\textit{(i)} the dark quark mass $m$,
\textit{(ii)} ``quadratic divergence'' from the $SU(2)_W$ gauge interaction,
and 
\textit{(iii)} a finite correction from electroweak symmetry breaking.

Firstly, all of the dark pions have the common squared mass originating from the dark quark mass. Eq.~(\ref{eq:meson potential}) contains mass terms of them. It is given by
\begin{align}
  m_{\Pi}^2 = \frac{4mv_d^3}{f_d^2}\cos\frac{\theta}{3}.\label{eq:dquarkmass}
\end{align}
We estimate $v_d$ using the large $N_c$ scaling~\cite{Manohar:1998xv},
\begin{align}
  v_d^3 \sim v_{\rm QCD}^3 \frac{f_d^3}{f_\pi^3} \sqrt{\frac{3}{N_c}},\label{eq:vev}
\end{align}
where $f_\pi$ is the decay constant of the QCD pion, and $v_{\text{QCD}}^3$ is the quark condensation $\langle \bar q q \rangle$ in the QCD
and is related to the squared mass of the QCD pion given by
\begin{align}
  m_{\pi,\text{QCD}}^2 = \frac{2(m_u+m_d)v_{\text{QCD}}^3}{f_{\pi}^2}.
\end{align}
Using these relations, we estimate $m_{\Pi}^2$ as
\begin{align}
  m_{\Pi}^2
 &= \frac{2 m_{\pi,\text{QCD}}^2}{(m_u + m_d)f_\pi} \sqrt{\frac{3}{N_c}} m f_d \cos\frac{\theta}{3} \nonumber\\
  &\simeq (7.6\,\text{TeV})^2\,\sqrt{\frac{3}{N_c}}\left(\frac{m}{1\,\text{TeV}}\right)\left(\frac{f_d}{1\,\text{TeV}}\right)\cos\frac{\theta}{3}.
\end{align}
Here we have used the value of the pion decay constant $f_{\pi}\simeq 93$ MeV and the pion mass $m_{\pi,\text{QCD}}\simeq 134.9$ MeV in the real QCD. We also took quark masses $m_u\simeq 2.16$ MeV and $m_d\simeq 4.67$ MeV \cite{Workman:2022ynf}.

\begin{figure}[tbp]
  \centering
  \begin{minipage}[b]{0.5\columnwidth}
    \begin{tikzpicture}
      \begin{feynhand}
        \vertex (pi1) at (-3,0) {$\Pi$};
        \vertex (pi2) at (3,0) {$\Pi$};
        \vertex (a) at (-1,0);
        \vertex (b) at (1,0);
        \propag [sca] (pi1) to (a);
        \propag [sca] (a) to (b);
        \propag [sca] (b) to (pi2);
        \propag [bos] (a) to [out=90,in=90, edge label=$W$] (b);
      \end{feynhand}
    \end{tikzpicture}
  \end{minipage}
  \begin{minipage}[b]{0.5\columnwidth}
    \begin{tikzpicture}[scale=1.5]
        \begin{feynhand}
            \vertex (x1) at (-2,0) {$\Pi$};
            \vertex (p) at (0,0);
            \vertex (q) at (0,1);
            \vertex (x2) at (2,0) {$\Pi$};
            \propag[sca] (x1) to (p);
            \propag[sca] (p) to (x2);
            \propag[bos] (p) to [out=150, in=180, edge label=$W$] (q); 
            \propag[bos] (q) to [out=0, in=30] (p);
        \end{feynhand}
    \end{tikzpicture}
  \end{minipage} 
  \caption{Feynman diagrams for radiative corrections from the $SU(2)_W$ gauge interaction to the dark pion masses.}
  \label{fig:wexc}
  %\end{center}
\end{figure}
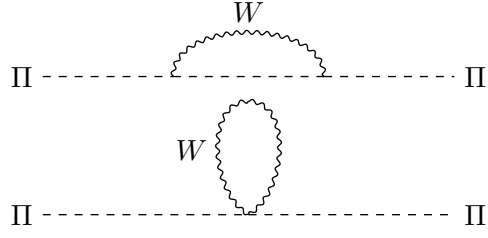
Secondly, we discuss the contributions from the $SU(2)_W$ gauge interaction.
The $SU(2)_W$ gauge interaction generates mass terms of $\chi$ and $\pi$ because
it explicitly breaks $SU(3)_L \times SU(3)_R$ symmetry. Furthermore, the $SU(2)_W$ gauge interaction does not respect $SU(3)_V$ symmetry, and the mass degeneracy between $\chi$ and $\pi$ is broken. The leading contribution comes from one-loop diagrams shown in figure~\ref{fig:wexc}.
%%%%%%%%%%%%%%%%%%%%%%%%%%%%%%%%%%%%%%%%%%%%%%%%%
This effect can be understood as quadratic UV divergence in the effective field theory, and this is the same way that electromagnetic interaction feeds the mass of the charged pion in the real QCD \cite{Das:1967it}.
The mass correction to the dark pions can be estimated by 
\cite{Farhi:1980xs, Dobrescu:1996jp}
\begin{align}
  \delta m_R^2 \sim C^2(R) \frac{\alpha_W(\Lambda_d)}{\alpha(\Lambda_{\text{QCD}})} \frac{\Lambda_d^2}{\Lambda_{\text{QCD}}^2} \left( m_{\pi_{\pm},\text{QCD}}^2-m_{\pi_0,\text{QCD}}^2 \right). \label{eq:qcd pion mass splitting} 
\end{align}
where $\alpha$ is the electromagnetic coupling and we took $\alpha(\Lambda_{\rm QCD}) \simeq 1/137$.
For $\chi$ and $\pi$, the quadratic Casimir is given as $C^2(\bm{3})=2$ and $C^2(\bm{5})=6$.
We estimate $\Lambda_d$ by utilizing the large $N_c$ scaling as $\Lambda_d/\Lambda_{\text{QCD}}\sim \sqrt{3/N_c}\,f_d/f_{\pi}$ and evaluate $\delta m_R^2$ by using eq. (\ref{eq:qcd pion mass splitting}) as 
\begin{align}
  \delta m_{\chi}^2 &\simeq2\Delta\simeq (1.1\,\text{TeV})^2\frac{3}{N_c}\left(\frac{f_d}{1\,\text{TeV}}\right)^2, \label{eq:delmchi}\\
  \delta m_{\pi}^2 &\simeq6\Delta\simeq (1.9\,\text{TeV})^2\frac{3}{N_c}\left(\frac{f_d}{1\,\text{TeV}}\right)^2,\label{eq:delmpi}\\
  \Delta &\equiv\frac{3}{N_c}\frac{\alpha_W(\Lambda_d)}{\alpha(\Lambda_{\text{QCD}})}\frac{f_d^2}{f_{\pi}^2}(m_{\pi^{\pm},\text{QCD}}^2-m_{\pi^0,\text{QCD}}^2).
\end{align}
Here we have used $m_{\pi_{\pm},\text{QCD}}\simeq 139.5$ MeV and $m_{\pi_0,\text{QCD}}\simeq 134.9$ MeV~\cite{Workman:2022ynf}. 
We evaluate $\alpha_W(\Lambda_d)$ by using the renormalization group equation for the $SU(2)_W$ gauge coupling at the one-loop level (see, e.g., ref.~\cite{Martin:1997ns}), 
\begin{align}
  \alpha_W(\Lambda_d)^{-1} \simeq \alpha_W(m_Z)^{-1}+\frac{19}{12\pi}\ln(\frac{\Lambda_d}{m_Z}),
\end{align}
where $\alpha_W(m_Z)\simeq 1/29$. Here we have used $\alpha(m_Z) \simeq 1/128$ and $\sin^2 \theta_W(m_Z) \simeq 0.23$ \cite{Workman:2022ynf}.
Typically, we obtain $\alpha_W(\Lambda_d=10\,\text{TeV})\simeq 1/32$. 

Lastly, we discuss a finite correction from electroweak symmetry breaking.
It induces a finite mass splitting among components of each multiplet. This effect can be understood as the Coulomb energy of the electric field of the electroweak gauge bosons around the electroweakly charged particle and it is not suppressed by the mass of the particles \cite{Cirelli:2005uq}. In a multiplet with $Y=0$, the mass splitting between a component with electric charge $Q$ and a neutral component is given by \cite{Cheng:1998hc, Feng:1999fu, Gherghetta:1999sw}\footnote{Two-loop results have been reported in ref.~\cite{Yamada:2009ve, Ibe:2012sx}.}
\begin{align}
  m_Q-m_0 &\simeq Q^2\Delta m, \\
  \Delta m &\equiv \alpha_Wm_W\sin^2\frac{\theta_W}{2}\simeq 166~\text{MeV}, \label{eq:Delta_m}
\end{align}
in the large mass limit $m_Q,~m_0 \gg m_W$.

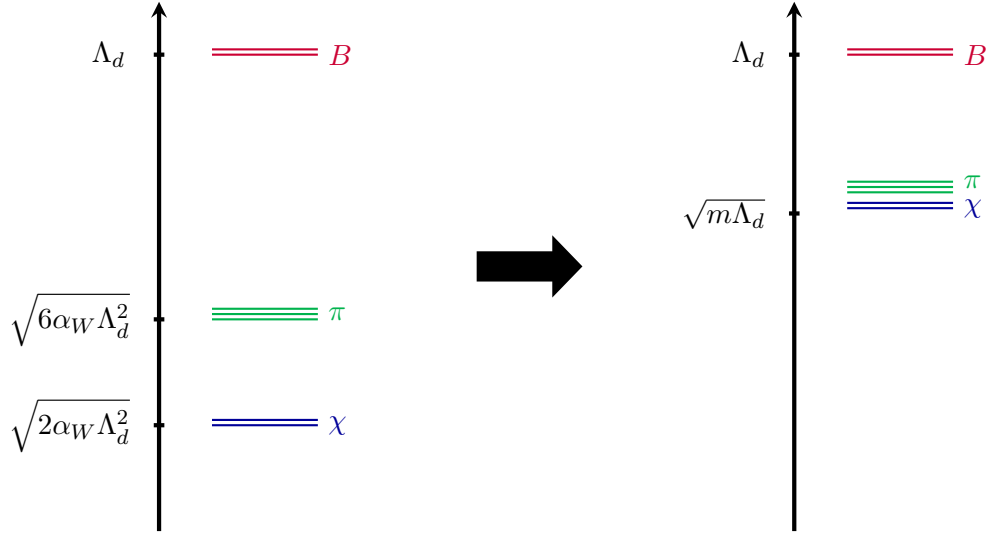
\begin{figure}[tbp]
  \centering
  \begin{tikzpicture}[scale=0.7]
    \draw[->,>=stealth, ultra thick](-7,-5)--(-7,5);
    \draw[ultra thick] (-7.1,-3)--(-6.9,-3) node[left=0.3cm] {$\sqrt{2\alpha_W\Lambda_d^2}$};
    \draw[ultra thick] (-7.1,-1)--(-6.9,-1) node[left=0.3cm]{$\sqrt{6\alpha_W\Lambda_d^2}$};
    \draw[ultra thick] (-7.1,4)--(-6.9,4) node[left=0.3cm]{$\Lambda_d\,$};
    \draw[thick, red!80!blue] (-6,4)--(-4,4) node[right]{$B$};
    \draw[thick, red!80!blue] (-6,4.1)--(-4,4.1);
    \draw[thick, blue!30!green] (-6,-1)--(-4,-1) ;
    \draw[thick, blue!30!green] (-6,-0.9)--(-4,-0.9)node[right]{$\pi$};
    \draw[thick, blue!30!green] (-6,-0.8)--(-4,-0.8);
    \draw[thick, blue!60!black] (-6, -3)--(-4,-3) node[right]{$\chi$};
    \draw[thick, blue!60!black] (-6,-2.9)--(-4, -2.9);
    \draw[->,>={Triangle[width=8mm, length=4mm]}, line width=4mm] (-1,0)--(1,0);
    \draw[->,>=stealth, ultra thick](5,-5)--(5,5);
    \draw[ultra thick] (4.9,4)--(5.1,4) node[left=0.3cm]{$\Lambda_d$};
    \draw[ultra thick] (4.9,1)--(5.1,1) node[left=0.3cm]{$\sqrt{m\Lambda_d}$};
    \draw[thick, red!80!blue] (6,4)--(8,4) node[right]{$B$};
    \draw[thick, red!80!blue] (6,4.1)--(8,4.1);
    \draw[thick,blue!60!black] (6,1.1)--(8,1.1)node[right]{$\chi$};
    \draw[thick,blue!60!black] (6,1.2)--(8,1.2);
    \draw[thick, blue!30!green] (6,1.4)--(8,1.4);
    \draw[thick, blue!30!green] (6,1.5)--(8,1.5);
    \draw[thick, blue!30!green] (6,1.6)--(8,1.6)node[right]{$\pi$};
\end{tikzpicture}
  \caption{Mass spectrum of the dark pions and the dark baryons. The mass spectrum in the limit of $m=0$ is shown in the left panel. Nonzero dark quark mass causes mass degeneracy of $\chi$ and $\pi$ as shown in the right panel.}
  \label{fig:mass spectrum}
\end{figure}

Combining all the corrections, we finally obtain the mass spectrum of each component as follows:
\begin{align}
  m_{\chi^0} &= m_{\chi},\\
  m_{\chi^{\pm}} &= m_{\chi}+\Delta m,\\
  m_{\pi^0} &= m_{\pi},\\
  m_{\pi^{\pm}} &= m_{\pi}+\Delta m,\\
  m_{\pi^{\pm\pm}} &= m_{\pi}+ 4\Delta m,
\end{align}
where
\begin{align}
  m_{\chi}^2 &= m_{\Pi}^2+\delta m_{\chi}^2,\label{eq:m_chi2}\\
  m_{\pi}^2 &= m_{\Pi}^2+\delta m_{\pi}^2.\label{eq:m_pi2}
\end{align} 
As we will see in section~\ref{sec:phenomenology}, the typical value of $f_d$ is $\gtrsim {\cal O}(1)~{\rm TeV}$. Thus, the finite mass splitting $\Delta m$ is always smaller than ``quadratic divergence'' $\delta m_{\chi}$ and $\delta m_{\pi}$, i.e., $\delta m_\pi > \delta m_\chi \gg \Delta m$ is hold.
Note that if the dark quark mass becomes larger, the masses of $\chi$ and $\pi$ are approximately degenerated because of $m_\Pi \gg \delta m_\chi,~\delta m_\pi$. See also figure~\ref{fig:mass spectrum}. The approximate degeneracy affects the dynamics of the DM candidate, as we will see in the next section.
\subsection{Heavier composite particles}
Before closing this section, let us briefly comment on heavy composite particles whose mass is of the order of $\Lambda_d \sim 4\pi f_d$.
We utilize the chiral Lagrangian eq.~(\ref{eq:Leff}) to evaluate the relic abundance of $\chi$ in the next section.
Since this chiral Lagrangian is an effective Lagrangian with cutoff scale $\sim\Lambda_d$, the effects of heavy composite particles are not included.

$\rho$ meson-like spin 1 resonance is one of such heavy particles. As the real QCD, dark $\rho$ meson couples with the dark pions and it can affect the annihilation of the dark pions. This means that our analysis based on the chiral Lagrangian is reliable if the mass of $\rho$ meson is sufficiently larger than the mass of pions, i.e., $m\lesssim \Lambda_d$. This implies a limitation of our analysis.

The dark baryon is also a heavy composite particle whose mass is $\sim \Lambda_d$. As we have already discussed, the lightest baryon $B$ is also stable due to the $U(1)_{\Psi}$ global symmetry and is also a DM candidate. We do not rely on the chiral Lagrangian for the estimation of the relic abundance of $B$, but utilize the naive scaling of the real QCD in section \ref{sec:DM abundance}.

\section{Numerical analysis} \label{sec:phenomenology}
In this section, we show the results of our numerical analysis of available parameter space in the current model. For concreteness, we focus on a specific case where $N_c = 3$.
We calculate the annihilation cross section of the DM candidates and the electron EDM, and we show the parameter space to be consistent with $\Omega_{\rm DM} h^2 = 0.12$.

\subsection{Relic abundance of DM} \label{sec:DM abundance}
As we discussed in the previous section, the lightest $G$-parity odd dark pion $\chi$ and the lightest dark baryon $B$ are stable particles and are the DM candidates. For $N_c=3$, $B$ is a spin $1/2$ fermion, and octet in the $SU(3)_V$ symmetry if we neglect the $SU(2)_W$ gauge coupling. Nonzero gauge coupling breaks the $SU(3)_V$ symmetry explicitly and this octet baryon is split into triplet and quintuplet in the $SU(2)_W$ symmetry. Similar to the dark pion, a neutral component in the triplet baryon becomes the lightest baryon because of the $SU(2)_W$ radiative corrections. Their number densities are determined by the freeze-out mechanism. In this subsection, we discuss how we estimate the relic abundance of $\chi$ and $B$.

\subsubsection{Relic abundance of the dark pions}
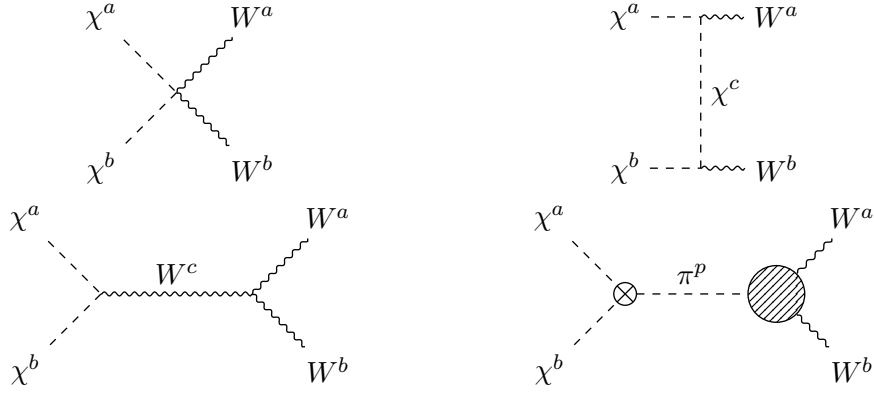
\begin{figure}[tbp]
  \centering
  \begin{minipage}[h]{0.45\linewidth}
    \centering
    \begin{tikzpicture}
      \begin{feynhand}
        \vertex (x1) at (-1,1) {$\chi^a$};
        \vertex (x2) at (-1,-1) {$\chi^b$};
        \vertex (w1) at (1,1) {$W^a$};
        \vertex (w2) at (1,-1) {$W^b$};
        \vertex (o) at (0,0);
        \propag [sca] (x1) to (o);
        \propag [sca] (x2) to (o);
        \propag [bos] (o) to (w1);
        \propag [bos] (o) to (w2);
      \end{feynhand} 
    \end{tikzpicture}
  \end{minipage}
  \begin{minipage}[h]{0.45\linewidth}
    \centering
    \begin{tikzpicture}[scale=1.0]
      \begin{feynhand}
        \vertex (x1) at (-1,1) {$\chi^a$};
        \vertex (x2) at (-1,-1) {$\chi^b$};
        \vertex (p) at (0,1);
        \vertex (q) at (0,-1);
        \vertex (w1) at (1,1) {$W^a$};
        \vertex (w2) at (1,-1) {$W^b$};
        \propag [sca] (x1) to (p);
        \propag [sca] (p) to [edge label=$\chi^c$] (q);
        \propag [bos] (p) to (w1);
        \propag [sca] (x2) to (q);
        \propag [bos] (q) to (w2);
      \end{feynhand}
    \end{tikzpicture}
  \end{minipage}
  \begin{minipage}[h]{0.45\linewidth}
    \centering
    \begin{tikzpicture}
        \begin{feynhand}
            \vertex (x1) at (-2,1) {$\chi^a$};
            \vertex (x2) at (-2,-1) {$\chi^b$};
            \vertex[crossdot] (p) at (-1,0);
            \vertex[NEblob] (q) at (1,0);
            \vertex (w1) at (2,1) {$W^a$};
            \vertex (w2) at (2,-1) {$W^b$};
            \propag [sca] (x1) to (p);
            \propag [bos] (p) to [edge label=$W^c$] (q);
            \propag [sca] (x2) to (p);
            \propag [bos] (q) to (w1);
            \propag [bos] (q) to (w2);
        \end{feynhand}  
    \end{tikzpicture}
  \end{minipage}
  \begin{minipage}[h]{0.45\linewidth}
    \centering
    \begin{tikzpicture}
        \begin{feynhand}
            \vertex (x1) at (-2,1) {$\chi^a$};
            \vertex (x2) at (-2,-1) {$\chi^b$};
            \vertex[crossdot] (p) at (-1,0){};
            \vertex[NEblob] (q) at (1,0){};
            \vertex (w1) at (2,1) {$W^a$};
            \vertex (w2) at (2,-1) {$W^b$};
            \propag [sca] (x1) to (p);
            \propag [sca] (p) to [edge label=$\pi^p$] (q);
            \propag [sca] (x2) to (p);
            \propag [bos] (q) to (w1);
            \propag [bos] (q) to (w2);
        \end{feynhand}  
    \end{tikzpicture}
  \end{minipage}
  \caption{Feynman diagrams for $\chi\chi\to WW$. $a$, $b$, and $c$ are indices of the adjoint representation of $SU(2)_W$. $p$ is a index of $\bm{5}$ representation of $SU(2)_W$. The crossed point indicates CP-violating vertex. The blob represents the WZW interactions.}
  \label{fig:to WW}
\end{figure}
Let us first discuss the relic abundance of the dark pion $\chi$.
Here we show the qualitative behavior of the DM annihilation cross section and the relic abundance of $\chi$. 
In addition to the $\chi^0\chi^0$ annihilation, coannihilation among $\chi^0$ and $\chi^\pm$ plays an important role because of the small mass splitting between $\chi^0$ and $\chi^\pm$. We denote $n_\chi$ as the total number density of $\chi^0$ and $\chi^\pm$. 
The time evolution of $n_\chi$ is described by the following Boltzmann equation:
\begin{align}
  \dot{n}_{\chi}+3Hn_{\chi} \simeq &- \left( \langle\sigma v\rangle_{WW} + \langle\sigma v\rangle_{\pi W} + \langle\sigma v\rangle_{\pi\pi} \right) \left(n_{\chi}^2-n_{\chi,eq}^{2}\right),\label{eq:boltzmanneq}
\end{align}
where $\langle\sigma v\rangle_{WW}$, $\langle\sigma v\rangle_{\pi W}$ and $\langle\sigma v\rangle_{\pi\pi}$ are the thermal averaged cross sections of $\chi\chi\to WW$, $\chi\chi\to\pi W$ and $\chi\chi\rightarrow\pi\pi$ respectively, and where $W$ represents the $SU(2)_W$ gauge bosons. Tree level diagrams which contribute to $\langle\sigma v\rangle_{WW}$, $\langle\sigma v\rangle_{\pi W}$ and $\langle\sigma v\rangle_{\pi\pi}$ are shown in figures~\ref{fig:to WW}, \ref{fig:pi V}, and \ref{fig:forbidann}, respectively. 
Note that annihilation processes into a pair of SM fermions and $Wh$ are $p$-wave suppressed and their effect is negligible compared to $\chi\chi\to WW$.
$\pi$ decays into two gauge bosons via WZW interaction and its decay width is $\Gamma_\pi \sim (\alpha_W/4\pi)^2 m_\pi^3 / f_d^2$. Since this $\Gamma_\pi$ is large enough compared to the Hubble rate $H$ at the freeze-out time of $\chi$, $\pi$ is in equilibrium with the thermal bath because of its decay and inverse decay processes. 

Let us discuss which annihilation process has the dominant effect in the Boltzmann equation.
Rough behavior of the annihilation cross section $\langle\sigma v\rangle_{WW}$, $\langle\sigma v\rangle_{\pi W}$, and $\langle\sigma v\rangle_{\pi\pi}$ are given by 
\begin{align}
  \langle\sigma v\rangle_{WW} &= \langle\sigma v\rangle_{WW}^{\rm MDM} + \langle\sigma v\rangle_{WW}^{\rm WZW},\label{eq:to WW}\\
  \langle\sigma v\rangle_{\pi W} &\propto \frac{g^2}{m_{\chi}^2}\left(\frac{m\sin(\theta/3)}{m_\chi}\right)^2,\label{eq:to piV}\\
  \langle\sigma v\rangle_{\pi\pi} &\propto \frac{1}{m_{\chi}^2}\frac{m_{\chi}^4}{f_d^4}\exp[-\frac{2(m_{\pi}-m_{\chi})}{T}], \label{eq:to pi pi}
\end{align}
where $T$ is the temperature of the thermal bath,\footnote{In the evaluation of
eqs.~\eqref{eq:to WW}--\eqref{eq:to pi pi}, we assume the mass difference in the $SU(2)_W$ multiplets, such as $m_{\pi^\pm}-m_{\pi^0}$, is much smaller than $T$.
}
$\langle\sigma v\rangle_{WW}^{\rm MDM}$ is the annihilation cross section of the minimal dark matter (MDM) calculated in refs.~\cite{Cirelli:2005uq, Cirelli:2007xd}, and $\langle\sigma v\rangle_{WW}^{\rm WZW}$ is the annihilation cross section which contains the WZW interactions (see bottom right panel of figure~\ref{fig:to WW}). They are given as
\begin{align}
\langle\sigma v\rangle_{WW}^{\rm MDM} \propto \frac{g^4}{m_\chi^2},\quad \langle\sigma v\rangle_{WW}^{\rm WZW}\propto \frac{g^4}{m_\chi^2}\left(\frac{m\sin(\theta/3)}{f_d}\right)^2\frac{m_\chi^4}{m_\pi^4},
\end{align}
In $\langle \sigma v \rangle_{WW}$, there is no interference between diagrams with and without the CP-violating interaction because of the helicities of the $SU(2)_W$ gauge bosons in the final state.
In $\langle \sigma v \rangle_{\pi W}$, we have neglected contributions from diagrams with WZW interaction because they suffer from $p$-wave suppression.
$\langle\sigma v\rangle_{\pi\pi}$ has the Boltzmann suppression factor because of the mass difference between $\chi$ and $\pi$.

For $m\simeq0$, the $SU(2)_W$ gauge interaction is the dominant source of the mass of the dark pions, and $\pi$ is $\sqrt{3}$ times heavier than $\chi$ as we can read from eqs.~(\ref{eq:delmchi}) and (\ref{eq:delmpi}). This mass difference results in the exponential suppression of $\langle\sigma v\rangle_{\pi\pi}$ during the freeze-out. Also, $\langle\sigma v\rangle_{\pi W}$ is negligible because it is suppressed by the dark quark mass $m$.
Thus, $\langle\sigma v\rangle_{WW}$ is dominant in the RHS of eq.~(\ref{eq:boltzmanneq}), and $\chi$ mainly annihilates into the electroweak gauge bosons. In this case, $\pi$ does not play a significant role in determining the relic abundance of $\chi$. Because $\chi$ is a triplet of $SU(2)_W$, the situation is the same as the MDM scenario \cite{Cirelli:2005uq, Cirelli:2007xd}. The relic abundance of $\chi$ in this regime is determined by $m_\chi$ and we obtain $\Omega_\chi h^2 = 0.12$ with $m_\chi \simeq 1.8~{\rm TeV}$ from the tree-level annihilation cross section\footnote{This analysis does not include the Sommerfeld effect. If we take it into account, the annihilation cross section of $\chi$ is enhanced and $m_{\chi}$ becomes as large as $m_{\chi}\simeq 2.5$~TeV \cite{Cirelli:2007xd}.}.

For the large $m$ regime, $m_{\Pi}^2$ given in eq.~\eqref{eq:dquarkmass} is the dominant contribution to the mass of the dark pions. In contrast to the $SU(2)_W$ gauge interaction, $m_\chi$ and $m_\pi$ have the same dependence on $m_\Pi^2$, and thus the mass ratio  $m_\pi / m_\chi$ gradually becomes close to 1 as the value of $m$ increases. For $m_\chi \simeq m_\pi$, $\langle\sigma v\rangle_{\pi\pi}$ becomes sizable because the Boltzmann factor $\exp\left( -2(m_{\pi}-m_{\chi})/T \right)$ is ${\cal O}(1)$ and it does not suppress the annihilation rate any more. Furthermore, the ratio between $\langle\sigma v\rangle_{WW}$ and $\langle\sigma v\rangle_{\pi\pi}$ scales as $\expval{\sigma v}_{WW}/\expval{\sigma v}_{\pi \pi} \sim g^4 f_d^4 / m_\chi^4 \sim (g^2 / 16\pi^2)^2 (4\pi f_d/m)^2$. Thus, for the large $m$ regime, $\chi \chi \to \pi \pi$ is the dominant annihilation channel, and our model naturally accommodates the forbidden dark matter scenario \cite{Griest:1990kh, DAgnolo:2015ujb}. 
\begin{figure}[tbp]
  \centering
  \begin{minipage}[h]{0.45\linewidth}
  \centering
  \begin{tikzpicture}[scale=1.0]
    \begin{feynhand}
      \vertex (x1) at (-2,1) {$\chi^a$};
      \vertex (x2) at (-2,-1) {$\chi^b$};
      \vertex[crossdot] (p) at (-1,0){};
      \vertex (q) at (1,0);
      \vertex (w1) at (2,1) {$\pi^p$};
      \vertex (w2) at (2,-1) {$W^b$};
      \propag [sca] (x1) to (p);
      \propag [sca] (p) to [edge label=$\pi^p$] (q);
      \propag [sca] (x2) to (p);
      \propag [sca] (q) to (w1);
      \propag [bos] (q) to (w2);
    \end{feynhand}
  \end{tikzpicture}
\end{minipage}
\begin{minipage}[h]{0.45\linewidth}
  \centering
  \begin{tikzpicture}[scale=1.0]
    \begin{feynhand}
      \vertex (x1) at (-1,1.5) {$\chi^a$};
      \vertex (x2) at (-1,-1.5) {$\chi^b$};
      \vertex[crossdot] (p) at (0,1){};
      \vertex (q) at (0,-1);
      \vertex (w1) at (1,1.5) {$\pi^p$};
      \vertex (w2) at (1,-1.5) {$W^a$};
      \propag [sca] (x1) to (p);
      \propag [sca] (p) to [edge label=$\chi^c$] (q);
      \propag [sca] (x2) to (q);
      \propag [sca] (p) to (w1);
      \propag [bos] (q) to (w2);
    \end{feynhand}
  \end{tikzpicture}
\end{minipage}
\begin{minipage}[h]{0.45\linewidth}
  \centering
  \begin{tikzpicture}[scale=1.0]
    \begin{feynhand}
      \vertex (x1) at (-2,1) {$\chi^a$};
      \vertex (x2) at (-2,-1) {$\chi^b$};
      \vertex (p) at (-1,0);
      \vertex[NEblob] (q) at (1,0){};
      \vertex (w1) at (2,1) {$\pi^p$};
      \vertex (w2) at (2,-1) {$W^b$};
      \propag [sca] (x1) to (p);
      \propag [bos] (p) to [edge label=$W^c$] (q);
      \propag [sca] (x2) to (p);
      \propag [sca] (q) to (w1);
      \propag [bos] (q) to (w2);
    \end{feynhand}
  \end{tikzpicture}
\end{minipage}
\begin{minipage}[h]{0.45\linewidth}
  \centering
  \begin{tikzpicture}[scale=1.0]
    \begin{feynhand}
      \vertex (x1) at (-1,1) {$\chi^a$};
      \vertex (x2) at (-1,-1) {$\chi^b$};
      \vertex (w1) at (1,1) {$\pi^p$};
      \vertex (w2) at (1,-1) {$W^b$};
      \vertex[NEblob] (o) at (0,0){};
      \propag [sca] (x1) to (o);
      \propag [sca] (x2) to (o);
      \propag [sca] (o) to (w1);
      \propag [bos] (o) to (w2);
    \end{feynhand} 
  \end{tikzpicture}
\end{minipage}
\caption{Feynman diagrams for $\chi\chi\to\pi W$. Indices are the same as ones given in figure~\ref{fig:to WW}.}
\label{fig:pi V}
\end{figure}
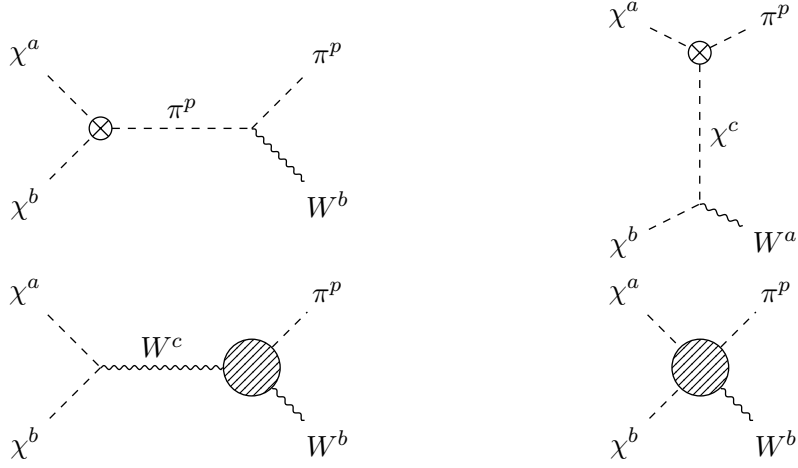
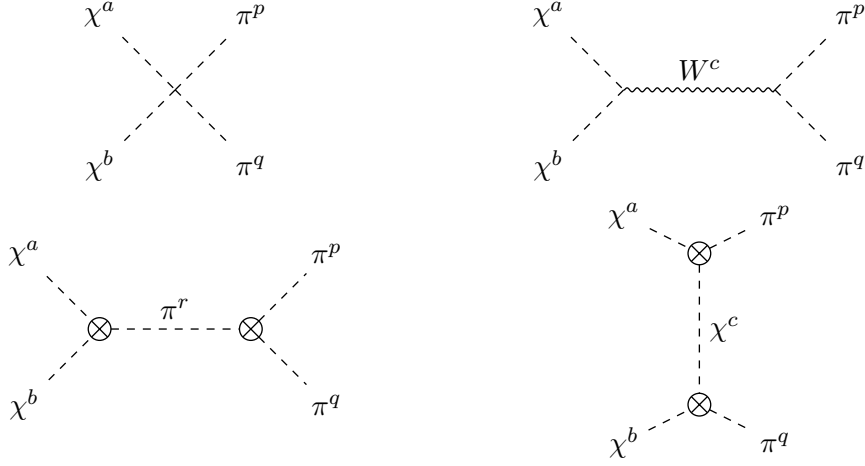
\begin{figure}[tbp]
  \centering
  \begin{minipage}[h]{0.45\linewidth}
  \centering
  \begin{tikzpicture}[scale=1.0]
    \begin{feynhand}
      \vertex (x1) at (-1,1) {$\chi^a$};
      \vertex (x2) at (-1,-1) {$\chi^b$};
      \vertex (w1) at (1,1) {$\pi^p$};
      \vertex (w2) at (1,-1) {$\pi^q$};
      \vertex (o) at (0,0);
      \propag [sca] (x1) to (o);
      \propag [sca] (x2) to (o);
      \propag [sca] (o) to (w1);
      \propag [sca] (o) to (w2);
    \end{feynhand} 
  \end{tikzpicture}
\end{minipage}
\begin{minipage}[h]{0.45\linewidth}
  \centering
  \begin{tikzpicture}[scale=1.0]
    \begin{feynhand}
      \vertex (x1) at (-2,1) {$\chi^a$};
      \vertex (x2) at (-2,-1) {$\chi^b$};
      \vertex (p) at (-1,0);
      \vertex (q) at (1,0);
      \vertex (w1) at (2,1) {$\pi^p$};
      \vertex (w2) at (2,-1) {$\pi^q$};
      \propag [sca] (x1) to (p);
      \propag [bos] (p) to [edge label=$W^c$] (q);
      \propag [sca] (x2) to (p);
      \propag [sca] (q) to (w1);
      \propag [sca] (q) to (w2);
    \end{feynhand}
  \end{tikzpicture}
\end{minipage}
\begin{minipage}[h]{0.45\linewidth}
  \centering
  \begin{tikzpicture}[scale=1.0]
    \begin{feynhand}
      \vertex (x1) at (-2,1) {$\chi^a$};
      \vertex (x2) at (-2,-1) {$\chi^b$};
      \vertex[crossdot] (p) at (-1,0){};
      \vertex[crossdot] (q) at (1,0){};
      \vertex (w1) at (2,1) {$\pi^p$};
      \vertex (w2) at (2,-1) {$\pi^q$};
      \propag [sca] (x1) to (p);
      \propag [sca] (p) to [edge label=$\pi^r$] (q);
      \propag [sca] (x2) to (p);
      \propag [sca] (q) to (w1);
      \propag [sca] (q) to (w2);
    \end{feynhand}
  \end{tikzpicture}
\end{minipage}
\begin{minipage}[h]{0.45\linewidth}
  \centering
  \begin{tikzpicture}[scale=1.0]
    \begin{feynhand}
      \vertex (x1) at (-1,1.5) {$\chi^a$};
      \vertex (x2) at (-1,-1.5) {$\chi^b$};
      \vertex[crossdot] (p) at (0,1){};
      \vertex[crossdot] (q) at (0,-1){};
      \vertex (w1) at (1,1.5) {$\pi^p$};
      \vertex (w2) at (1,-1.5) {$\pi^q$};
      \propag [sca] (x1) to (p);
      \propag [sca] (p) to [edge label=$\chi^c$] (q);
      \propag [sca] (x2) to (q);
      \propag [sca] (p) to (w1);
      \propag [sca] (q) to (w2);
    \end{feynhand}
  \end{tikzpicture}
\end{minipage}
\caption{Feynman diagrams for forbidden annihilation channels, $\chi\chi\to \pi\pi$. $p$, $q$, and $r$ are indices of $\bm{5}$ representation of $SU(2)_W$. }
\label{fig:forbidann}
\end{figure}

\subsubsection{Relic abundance of the dark baryon}
Let us estimate the annihilation cross section which is relevant to the relic abundance of the dark baryon from an analogy with the annihilation of the baryon in the real QCD.
In the non-relativistic limit, the $s$-wave annihilation cross section is parametrized as
\begin{align}
\sigma_B \simeq \frac{4\pi}{k}\Im a,
\end{align}
where $a$ is the scattering length, and $k$ is the center of mass momentum.
The annihilation cross section of baryon in the real QCD has been measured in OBELIX experiment \cite{OBELIX:1997nkw},
which shows $\Im a \simeq 0.5~{\rm fm} \simeq 2.4/m_N$ for $\bar n p$ annihilation where $m_N$ is the nucleon mass.
Then we estimate the thermal averaged annihilation cross section for the dark baryon as
\begin{align}
  \langle\sigma_B v\rangle \simeq c\frac{4\pi}{m_B^2}, \label{eq:baryon annihilation}
\end{align}
where $c$ is a ${\cal O}(1)$ parameter. Then, by using $\Omega h^2 \simeq 0.12 \times (2\times 10^{-26}~{\rm cm}^3/{\rm s} ) / \langle \sigma v\rangle $ \cite{Steigman:2012nb}, we estimate the relic abundance of the baryon as
\begin{align}
\Omega_B h^2 \simeq 0.12 \times \frac{1}{c} \left(\frac{f_d}{6.66\,\text{TeV}}\right)^2. \label{eq:omegah2 baryon}
\end{align}
Here we take a naive estimation of the baryon mass as $m_B\sim \Lambda_d\sim 4\pi f_d$.

\subsection{Electron electric dipole moment}
$\theta$ has physical CP-violating effect if $m \neq 0$ \cite{Draper:2016fsr, Redi:2016kip, Choi:2016hro, Draper:2018tmh}.
After integrating out the DM sector, the CP-violating effect is given by a dimension-six Weinberg operator with $SU(2)_W$ gauge field as
\begin{align}
{\cal L}_{\rm eff.} \ni C_{\tilde W} \epsilon^{abc} W_\mu^{a\nu} W_\nu^{b\lambda} \tilde W_\lambda^{c\mu}.
\end{align}
The Wilson coefficient $C_{\tilde W}$ can be estimated by NDA \cite{Choi:2016hro}:
\begin{align}
C_{\tilde W} \sim \frac{N_c}{(4\pi)^2}\frac{m\sin(\theta/3)}{(4\pi f_d)^3}\frac{g^3}{3}\times c_{\rm EDM},\label{eq:WeinbergOp from dark sector}
\end{align}
where $c_{\rm EDM}$ is a ${\cal O}(1)$ parameter. The one-loop diagram involving the Weinberg operator induces the finite size of the electron EDM \cite{Boudjema:1990dv, Gripaios:2013lea, Panico:2018hal, Kley:2021yhn}
 as
\begin{align}
d_e = \frac{5 g^2 m_e \sin\theta_W }{96\pi^2} C_{\tilde W}, \label{eq:EDM from WeinbergOp}
\end{align}
where $m_e$ is the electron mass. Here we used naive dimensional regularization for one-loop computation.\footnote{
Refs.~\cite{Panico:2018hal, Kley:2021yhn} obtained $d_e = (3 g^2 m_e\sin\theta_W / 32\pi^2) C_{\tilde W}$ from naive dimensional regularization and this is not consistent with eq.~(\ref{eq:EDM from WeinbergOp}).
In four-dimension, we can show that $\epsilon^{abc} W_\mu^{a\nu} W_\nu^{b\lambda} \tilde W_\lambda^{c\mu} = 3 [ W_\mu^{1\nu} W_\nu^{2\lambda} \tilde W_\lambda^{3\mu} - W_\mu^{2\nu} W_\nu^{1\lambda} \tilde W_\lambda^{3\mu}]$ by utilizing $\epsilon_{\mu\nu\rho\sigma} \epsilon^{\alpha\beta\rho\sigma} = -2(g^\alpha_\mu g^\beta_\nu - g^\alpha_\nu g^\beta_\mu)$, however, this does not hold in $d$ dimension.
We also confirmed that we can obtain $d_e = (g^2 m_e \sin\theta_W /32\pi^2) C_{\tilde W}$ from a part of Weinberg operator; ${\cal L}_{\rm eff.} \ni C_{\tilde W} [ W_\mu^{1\nu} W_\nu^{2\lambda} \tilde W_\lambda^{3\mu} - W_\mu^{2\nu} W_\nu^{1\lambda} \tilde W_\lambda^{3\mu}]$ as shown in ref.~\cite{Boudjema:1990dv}.
Thus, the result in refs.~\cite{Panico:2018hal, Kley:2021yhn} is obtained if we allow this transformation of the Weinberg operator \textit{before} analytic continuation to $d$-dimensional space.
If not, we obtain our eq.~(\ref{eq:EDM from WeinbergOp}).}

\subsection{Results}\label{sec:results}
In this section, we show the results of our numerical analysis. The current model has three free parameters; the decay constant $f_d$, the dark quark mass $m$, and $\theta$. We calculate the relic abundance of $\chi$, denoted as $\Omega_\chi h^2$, by using \texttt{micrOMEGAs} \cite{Belanger:2018ccd} and \texttt{FeynRules} \cite{Alloul:2013bka}.  For the relic abundance of $B$, denoted as $\Omega_B h^2$, we use eq.~(\ref{eq:omegah2 baryon}).
The total DM energy density is obtained as $\Omega_{\rm DM} h^2 = \Omega_\chi h^2 + \Omega_B h^2$. 

\subsubsection{CP-conserving case ($\theta=0$)}
\begin{figure}[tbp]
  \centering
    \includegraphics[width=0.8\hsize]{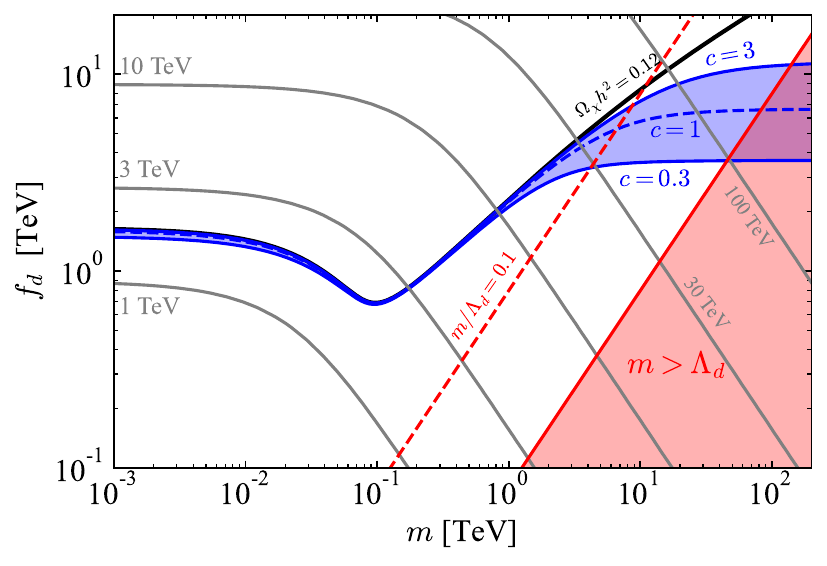}
    \caption{The parameters to give $\Omega_{\rm DM} h^2 = \Omega_{\chi}h^2 + \Omega_B h^2 = 0.12$ is shown by assuming $\theta = 0$ and $N_c=3$. $\Omega_{\chi} h^2 = 0.12$ is achieved on the black curve.
    $\Omega_{\rm DM} h^2 = 0.12$ is achieved on the blue band. For this blue band, we vary a coefficient $c$ in the baryon annihilation cross section eq.~(\ref{eq:baryon annihilation}) from 0.3 to 3. $c=1$ is indicated with the blue dotted curve. The gray contours indicate the mass of $\chi^0$. The red dashed line indicates $m/\Lambda_d=0.1$. In the red-filled region, we cannot trust our analysis based on the chiral Lagrangian because of $m > \Lambda_d$.}
  \label{fig:abundance}
\end{figure}
\begin{figure}[tbp]
  \centering
  \includegraphics[width=0.48\hsize]{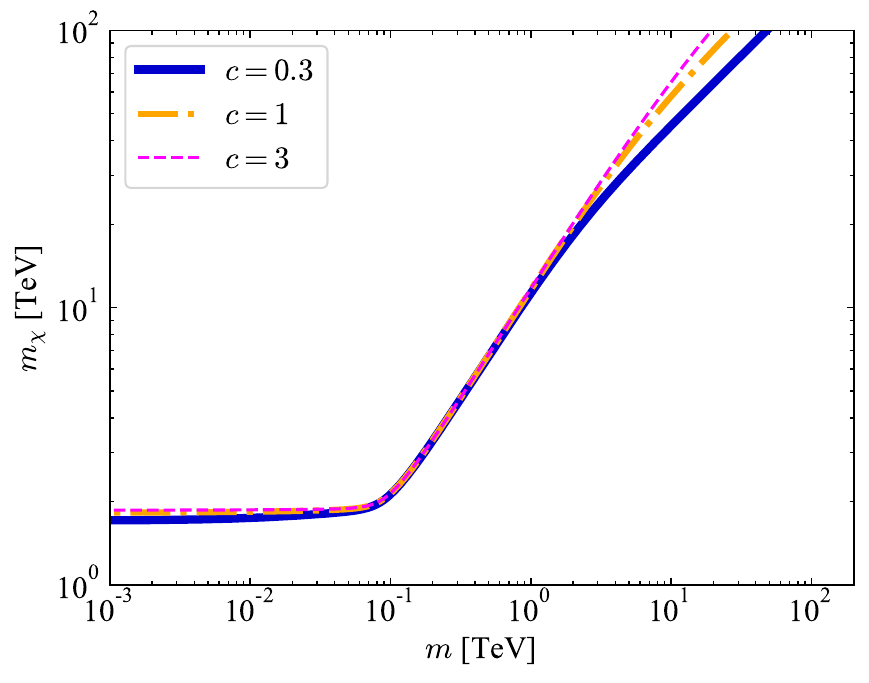}    \includegraphics[width=0.48\hsize]{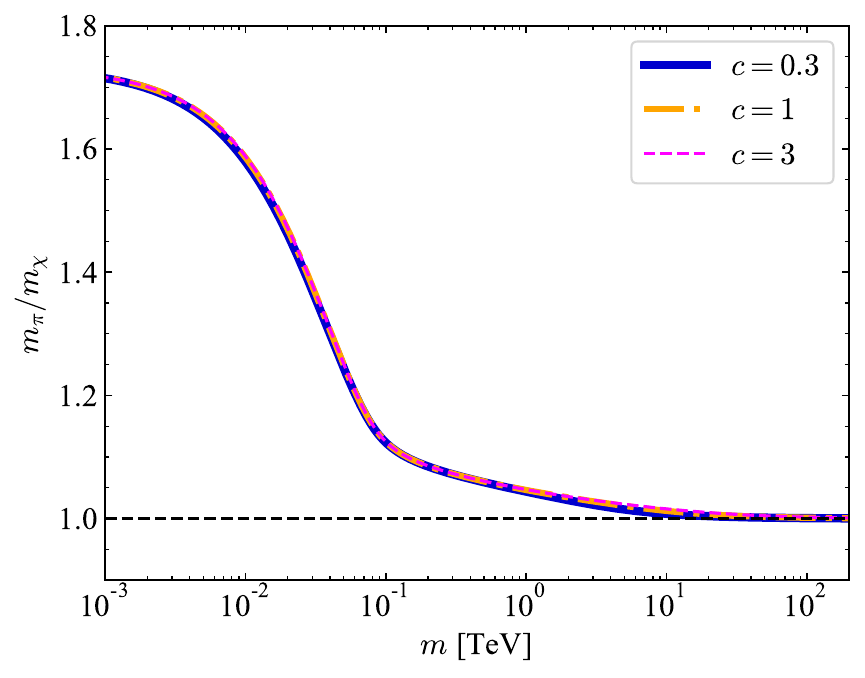}
  \includegraphics[width=0.48\hsize]{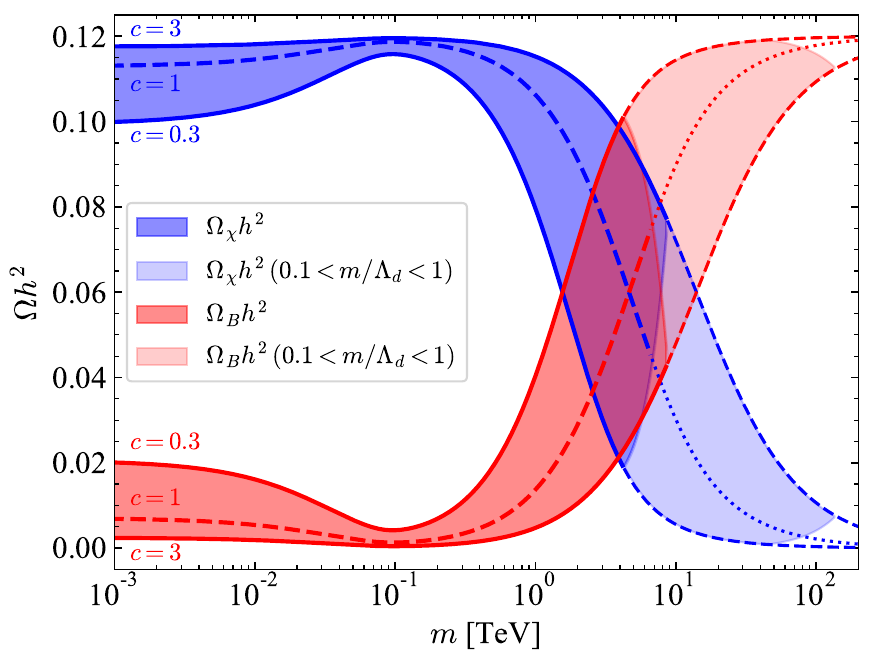}
  \caption{\emph{Upper left:} $m_\chi$ as a function of $m$ with $c=0.3$ (the blue solid line), $c=1$ (the orange dash-dotted line), and $c=3$ (the magenta dashed line). \emph{Upper right: }The mass ratio $m_{\pi}/m_{\chi}$ as a function of $m$ with $c=0.3$ (the blue solid line), $1$ (the orange dash-dotted line), and $3$ (the magenta dashed line). \emph{Lower:} $\Omega_\chi h^2$ and $\Omega_B h^2$ as functions of $m$ are shown as a blue band and a red band, respectively. The width of both bands comes from the uncertainties of the baryon annihilation cross section as indicated in eq.~(\ref{eq:baryon annihilation}). We show the regions where $0.1<m/\Lambda_d<1$ with lighter colors and the regions where $m>\Lambda_d$ are left white. In these three panels, we choose $f_d$ to achieve $\Omega_{\text{DM}}h^2 = 0.12$.} \label{fig:mass ratio and dm composition}
\end{figure}
First, we discuss a case with $\theta = 0$. As $f_d$ becomes larger with given $m$, the mass of dark pions and dark baryons becomes heavier, and their interaction becomes weaker. Thus, with given $m$, $\Omega_{\rm DM}h^2$ monotonously increases as a function of $f_d$. We find a value of $f_d$ for given $m$ to obtain the measured value of the DM energy density, $\Omega_{\rm DM}h^2 = 0.12$. In the blue-filled region of figure~\ref{fig:abundance}, the measured value of the DM energy density is explained by the freeze-out mechanism. Here, we vary the unknown $\mathcal{O}(1)$ parameter $c$ in eq.~(\ref{eq:omegah2 baryon}) as $0.3 \leq c \leq 3$. The effect on the determination of $f_d$ from the uncertainty on $c$ becomes larger if the fraction of the baryon DM $\Omega_B / \Omega_\chi$ is larger. By assuming $c=1$, we show $m_\chi$, $m_{\pi}/m_{\chi}$, $\Omega_{\chi}h^2$, and $\Omega_Bh^2$ as a function of $m$ in figure~\ref{fig:mass ratio and dm composition}.

For $m \lesssim 0.1~{\rm TeV}$, the main source of the masses of $\pi$ and $\chi$ is the radiative corrections from electroweak interaction and the properties of the dark sector are similar to ref.~\cite{Bai:2010qg}. Figure~\ref{fig:mass ratio and dm composition} shows that the mass splitting between $\chi$ and $\pi$ is large in this regime, and a pair of $\chi$ particles mainly annihilate into the electroweak gauge bosons. Since the dark baryon $B$ has a large annihilation cross section that saturates the unitarity bound, the dominant component of the DM is $\chi$. As a result, the uncertainty of the baryon annihilation cross section hardly affects the value of $f_d$ to be consistent with $\Omega_{\rm DM} h^2 = 0.12$ as shown in figure~\ref{fig:abundance}, and the relic abundance of $\chi$ for $m \lesssim 0.1~{\rm TeV}$ is determined by $m_\chi$. We find that $m_\chi \simeq 1.8$~TeV for $m \lesssim 0.1~{\rm TeV}$, and this value is almost independent of $m$. The fraction of $B$ in the DM is $\Omega_B / \Omega_\chi \sim {\cal O}(0.1)$.

For $m \gtrsim 0.1~{\rm TeV}$, the DM freeze-out process dramatically changes from the scenario in ref.~\cite{Bai:2010qg}. 
As we discussed in section \ref{sec:mass spectrum}, the dark quark mass universally contributes to both $m_\chi$ and $m_\pi$. The upper right panel of figure~\ref{fig:mass ratio and dm composition} shows that the mass ratio $m_\pi / m_\chi \lesssim 1.1$ for $m \gtrsim 0.1~{\rm TeV}$. Due to this degeneracy between $m_\pi$ and $m_\chi$, the Boltzmann factor does not suppress $\langle \sigma v \rangle_{\pi\pi}$, and the forbidden channels, $\chi \chi \to \pi \pi$, significantly contribute to the annihilation of $\chi$ in addition to the electroweak gauge boson channel, as discussed in section \ref{sec:DM abundance}. Thus, the total annihilation cross section is larger than the MDM case and the DM mass $m_\chi$ is larger than $1.8~{\rm TeV}$ to be consistent with $\Omega_{\rm DM} h^2 = 0.12$. This is the reason why the blue curve in figure~\ref{fig:abundance} indicates the heavier $m_\chi$ region as the value of $m$ increases for $m \gtrsim 0.1~{\rm TeV}$. As $m$ increases, the fraction of the baryon in the DM becomes larger as shown in the right panel of figure~\ref{fig:mass ratio and dm composition}.

For $m\gtrsim {\cal O}(1)~{\rm TeV}$, 
figure~\ref{fig:abundance} shows that $m_\chi \gtrsim 10$~TeV and $f_d \gtrsim 2$~TeV to obtain the right amount of the DM relic abundance. In this case, the mass of the dark pions is comparable to the mass of the baryon $m_B\sim 4\pi f_d$, and $\Omega_\chi \lesssim \Omega_B$ as shown in figure~\ref{fig:mass ratio and dm composition}. However, there is a large uncertainty in the calculation of the relic abundance. 
First, the baryon annihilation cross section has the unknown $\mathcal{O}(1)$ factor $c$. This $c$ brings the large uncertainty in the relic abundance of the baryon as can be seen in figures \ref{fig:abundance} and \ref{fig:mass ratio and dm composition}. 
Second, we implicitly assume that the contributions to the annihilation processes from heavier mesons, such as $\rho$, are negligible. 
This is a good approximation if the mass of $\rho$ mesons is much heavier than $m_\chi$ and $m_\pi$. 
However, for the heavier $m$ regime, the mass of the $\rho$ mesons is expected to be in the same order as $\chi$ and $\pi$.
These two are the sources of uncertainty in the calculation of the DM relic abundance. Our naive expectation is that the pion and baryon annihilation cross section saturates unitarity bound if their mass is around $m_B \sim m_\pi \sim {\cal O}(100)~{\rm TeV}$, and then $f_d (\sim m) \sim m_B/4\pi \sim {\cal O}(10)~{\rm TeV}$. 
If we take even larger $m \gg f_d$, we expect that our scenario is smoothly connected to the scenario discussed in ref.~\cite{Mitridate:2017oky} though our analysis based on chiral Lagrangian cannot be applied for the case with $m / \Lambda_d \gtrsim \mathcal{O}(1)$, which is shown by the red shaded region in figure~\ref{fig:abundance}. Thus, it is difficult to determine ${\cal O}(1)$ factor in both $\Omega_\chi h^2$ and $\Omega_B h^2$ with $m \sim \Lambda_d$.
We also show the line of $m / \Lambda_d  = 0.1$ in figure~\ref{fig:abundance} as a conservative limit of our analysis.
\subsubsection{CP-violating case ($\theta\neq0$)}
\begin{figure}[tbp]
  \centering
  \includegraphics[width=0.7\hsize]{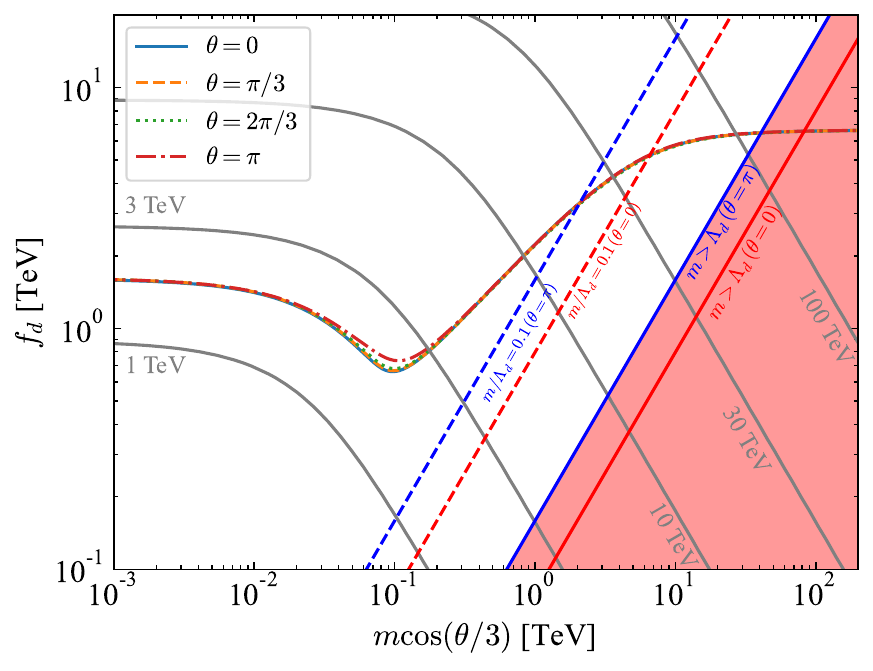}
  \caption{The parameter to give $\Omega_{\text{DM}}h^2=0.12$ is shown by assuming $\theta=0, \pi/3, 2\pi/3,$ and $\pi$. Note that we set the real part of the dark quark mass $m\cos(\theta/3)$ as the horizontal axis. The gray contours indicate the mass of $\chi^0$. In the red-filled region, we cannot trust our analysis based on the chiral perturbation because of $m>\Lambda_d$. Since the boundary depends on $\theta$ in this plane, we draw red and blue lines for $\theta=0$ and $\theta=\pi$ respectively.  The red- or blue-dashed lines indicate $m/\Lambda_d=0.1$.}\label{fig:with cp}
  ~\\[3mm]
  \includegraphics[width=0.7\hsize]{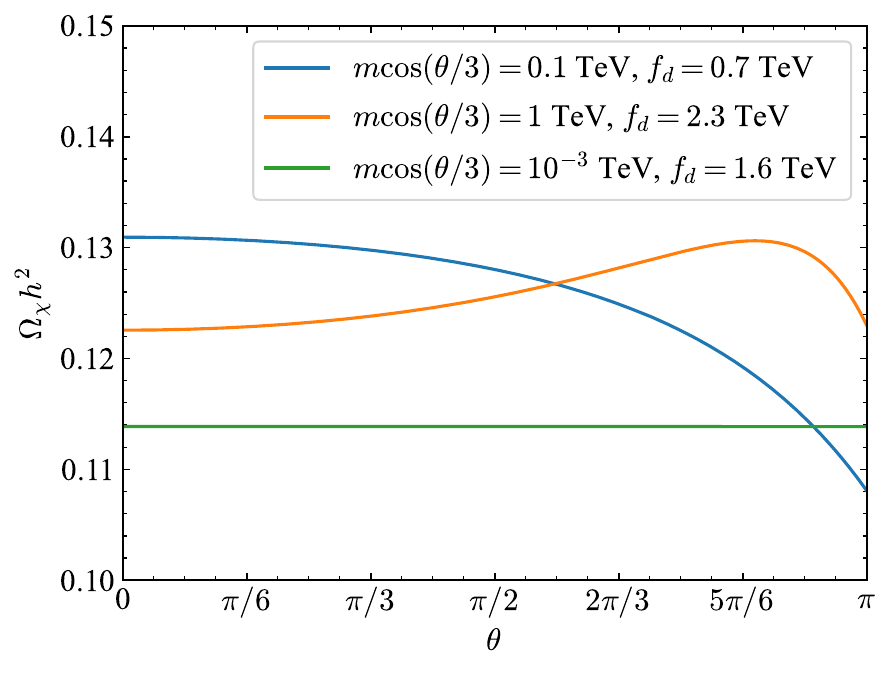}
  \caption{$\theta$-dependence in $\Omega_\chi h^2$ for given $m\cos(\theta/3)$ and $f_d$. We take $(m\cos(\theta/3),f_d) = (0.1~{\rm TeV},~0.7~{\rm TeV})$ for the blue curve, $(1~{\rm TeV},~2.3~{\rm TeV})$ for the orange curve, and $(10^{-3}~{\rm TeV},~1.6~{\rm TeV})$ for the green curve.}\label{fig:theta dependence}
\end{figure}

Next, we discuss a case with $\theta \neq 0$. 
In this case, CP-violating trilinear couplings, $\chi\chi\pi$ and $\pi\pi\pi$, are induced from eq.~\eqref{eq:meson potential}.
As we discussed in section~\ref{sec:models}, $m_{\chi}$ is determined by $m\cos(\theta/3)$ and $f_d$. Figure~\ref{fig:with cp} shows the value of $f_d$ that gives the measured value of the DM energy density for given $m\cos(\theta/3)$. Also, the impact of $\theta$ on the DM relic abundance is negligible in most cases and we can see a small effect at $m\cos(\theta/3) \sim 0.1~{\rm TeV}$.
Figure~\ref{fig:theta dependence} shows how $\Omega_\chi h^2$ depends on $\theta$ for given $m\cos(\theta/3)$ and $f_d$.
For $m\cos(\theta/3)\lesssim 0.1~{\rm TeV}$, the leading annihilation process is $\chi\chi\to WW$. 
The CP-violating trilinear couplings are proportional to $m\sin(\theta/3)$. 
For small $m\cos(\theta/3)$, the effect of the CP-violation on the determination of $\Omega_\chi h^2$ is negligible. In particular, the relic abundance is almost independent of $\theta$ for $m\cos(\theta/3) \lesssim 10^{-3}~{\rm TeV}$ as shown by the green curve in figure~\ref{fig:theta dependence}. 
For large $m\cos(\theta/3)$, the $\theta$-dependence in $\Omega_\chi h^2$ is visible and affects the cross section of $\chi\chi \to WW$. We can see its effect at $m\cos(\theta/3) \sim 0.1~{\rm TeV}$. See also the blue curve in figure~\ref{fig:theta dependence}. For $m\cos(\theta/3) \gtrsim 0.1~{\rm TeV}$, the relic abundance of $\chi$ is mainly determined by the cross section of the forbidden channel $\chi\chi\to\pi\pi$. As shown in the orange curve in figure~\ref{fig:theta dependence}, $\Omega_{\rm \chi}h^2$ changes at most 10\% by $\theta$ in this regime.
To summarize, the mass spectrum and the CP-conserving interaction determine the relic abundance in most cases, and the CP-violating interaction has only a little impact on this calculation.
\begin{figure}[tbp]
    \centering
    \includegraphics[width=0.48\hsize]{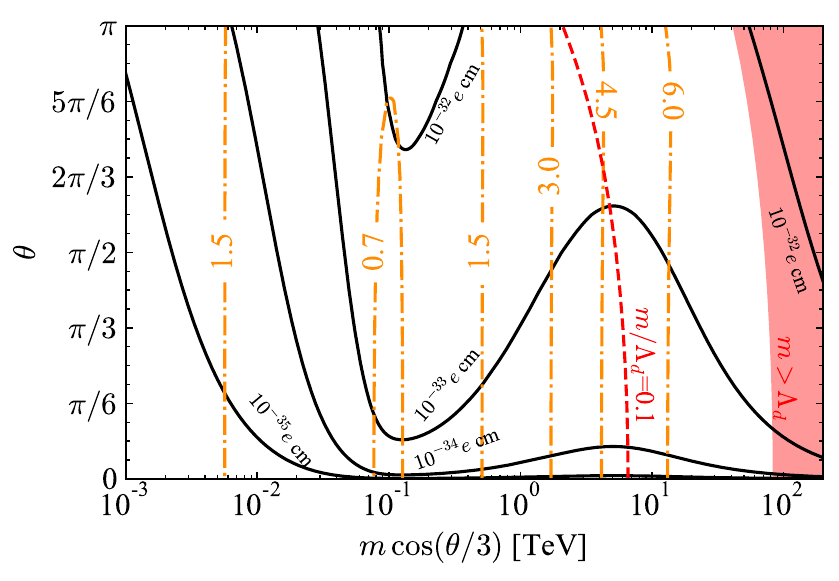}
    \includegraphics[width=0.48\hsize]{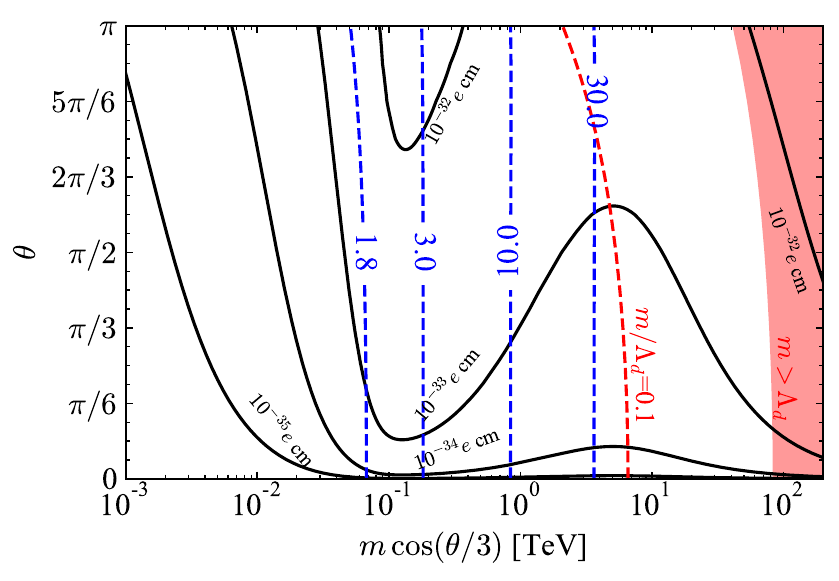}
  \caption{\emph{Left: }$f_d$ and $d_e$ in $m\cos(\theta/3)$ - $\theta$ plane. Orange dash-dotted lines indicate the value of $f_d$ in the unit of TeV. Black contours show the predicted value of the electron EDM.  \emph{Right: }$m_\chi$ and $d_e$ in $m\cos(\theta/3)$ - $\theta$ plane. Blue dashed lines indicate the value of $m_\chi$ in the unit of TeV. The horizontal axis is the real part of the dark quark mass $m\cos(\theta/3)$, and the vertical axis is $\theta$. At each point, $f_d$ is chosen to achieve $\Omega_{\rm DM} h^2=0.12$.
  Here we set the $\mathcal{O}(1)$ parameters in eq. (\ref{eq:baryon annihilation}) and (\ref{eq:WeinbergOp from dark sector}) as $c=1$ and $c_{\rm EDM}=1$, respectively.
  Red dashed lines indicate $m/\Lambda_d =0.1$.
  In the red-filled region, we cannot trust our analysis based on the chiral Lagrangian because of $m > \Lambda_d$. }
  \label{fig:m vs theta}
\end{figure}

In figure~\ref{fig:m vs theta}, we show the value of $f_d$ and the electron EDM in $m\cos(\theta/3)$-$\theta$ plane. At each point in this plane, $f_d$ is chosen to obtain $\Omega_{\rm DM} h^2 = 0.12$. We calculate the electron EDM by using eqs.~(\ref{eq:WeinbergOp from dark sector}) and (\ref{eq:EDM from WeinbergOp}) with $c_{\rm EDM}=1$. The maximal value of the electron EDM in the parameter space is $\sim 10^{-32}~e~{\rm cm}$.
The value of the electron EDM in this model is below the current upper bound, $|d_e| < 4.1\times 10^{-30} e~{\rm cm}$ \cite{Roussy:2022cmp}, and is out of the reach of the ACME III experiment which will probe the electron EDM at the level of $0.3\times 10^{-30} e~{\rm cm}$ \cite{Doyle:2016edm}. In order to probe the parameter space, further improvements in the electron EDM experiments are required. 

\section{Signals} \label{sec:signals}
In this section, we briefly comment on possible signals to probe the current DM model. The phenomenology of $SU(2)_W$ triplet real scalar DM has been discussed in, e.g., refs.~\cite{Cirelli:2005uq, Araki:2011hm, Abe:2014gua, Chiang:2020rcv, Arakawa:2021vih, Katayose:2021mew}. Some aspects of the dark pion DM $\chi$ are similar to the elementary scalar DM case. However, the current DM model has richer signals from other composite particles and strong dynamics of the dark sector.

\subsection{Direct detection}
The direct detection experiments can probe scattering between the DM and nucleons, and it is sensitive to scattering amplitude between the DM and quark/gluon. The leading contribution comes from one-loop diagrams for scattering with quarks and two-loop diagrams for scattering with gluons. Thus, spin-independent cross section $\sigma_{\rm SI}$ for scattering between $SU(2)_W$ triplet scalar $\chi$ and nucleons is suppressed by the loop factor. We are interested in the case with the DM mass to be $\gtrsim {\cal O}(1)~{\rm TeV}$ and $\sigma_{\rm SI}$ should be almost independent of the DM mass and its spin as shown in ref.~\cite{Chen:2023bwg}. 
Note that one-loop radiative correction induces a coupling between $\chi$ and the SM Higgs as ${\cal L} \ni -\lambda_{\chi H} |\chi|^2 |H|^2$ with $\lambda_{\chi H} \sim \alpha_W^2$. However, contribution to the spin-independent cross section $\sigma_{\rm SI}$ from this $\lambda_{\chi H}$ coupling is suppressed by $m_W / m_\chi$ \cite{Katayose:2021mew, Chen:2023bwg} compared to diagram without the Higgs propagator and its effect is small. Ref.~\cite{Hisano:2015rsa} shows $\sigma_{\rm SI} \simeq 2 \times 10^{-47}~{\rm cm}^2$ for wino DM with QCD NLO corrections. We expect $\chi$ should have a similar value of $\sigma_{\rm SI}$ and it is much below the current constraints by LZ \cite{LZ:2022lsv} and XENONnT \cite{XENON:2023cxc}. However, for $m_\chi \lesssim 4~{\rm TeV}$, $\sigma_{\rm SI}$ is still above the neutrino floor \cite{OHare:2021utq}, and we could have a chance to detect DM in the direct detection experiment in the future.

Since $\sigma_{\rm SI}$ is almost independent of the DM mass and its spin in the case with the DM mass to be $\gtrsim {\cal O}(1)~{\rm TeV}$, we can estimate that the spin-independent cross section of the dark baryon $B$ becomes similar to that of $\chi$. Furthermore, $B$ is subdominant for the DM abundance, and its mass $m_B$ is larger than the dark pion mass $m_\chi$. Thus, the number density of $B$ becomes smaller than $\chi$ and thus the signal rate of $B$ also becomes smaller. 

\subsection{Indirect detection}
The indirect detection experiments are one of the most promising ways to probe the current model. By assuming wino-like annihilation, the region with $m_\chi \lesssim 20~{\rm TeV}$ already started to be partly constrained by HESS \cite{HESS:2018cbt} and MAGIC \cite{MAGIC:2022acl} though these constraints suffer from the uncertainty of the density profile of our galaxy. In the analysis of indirect detection of $SU(2)_W$ charged DM, the Sommerfeld effect significantly changes its annihilation cross section \cite{Hisano:2002fk, Hisano:2003ec, Hisano:2004ds} because of the large mass hierarchy between $\chi$ and electroweak gauge bosons. Also, the Sommerfeld effect has an impact on the calculation of the relic abundance of the MDM \cite{Cirelli:2007xd}, and we expect a parameter to give the observed relic abundance in the current model to be affected by ${\cal O}(1)$ factor compared to our analysis presented in section~\ref{sec:phenomenology}, which used tree-level annihilation cross section. Compared to the MDM case, our model is more involved because the final state particle $\pi$ in a forbidden channel is also charged under $SU(2)_W$. We left a comprehensive analysis including the Sommerfeld effect for future study \cite{AbeSatoYamanakaSommerfeld}.

\subsection{Collider experiments}
The particles in the dark sector of the current model are heavier than $1~{\rm TeV}$, and they have only electroweak gauge interaction with the SM particles. Thus, it is difficult to probe these particles in the LHC experiment. However, it is interesting to discuss the possibility of probing this model at a future high energy collider such as the 100 TeV $pp$ collider. Since the DM $\chi$ has $SU(2)_W$ charge, it can be pair-produced via electroweak interaction, such as $W^{+*} \to \chi^0 \chi^+$. At 100 TeV $pp$ collider, $m_\chi = {\cal O}(1)~{\rm TeV}$ region could be probed depending on pileup scenario \cite{Chiang:2020rcv}. In addition to $\chi$, $G$-parity even $\pi$ would be an interesting target in future collider experiments. $\pi$ can be produced from, e.g., $V^* \to \pi V' ~(V,V'=\gamma,W,Z)$ via WZW interaction term, and $\rho^* \to \pi \pi$ where $\rho$ is spin-1 resonance in the dark sector \cite{Kilic:2009mi, Draper:2018tmh}. $\pi$ can decay into a pair of photons and these production channels result in a multi-photon channel. So we expect that $\pi$ provides a clean channel and this would be an interesting target in the collider experiments. Since 100 TeV $pp$ collider can have sensitivity on $Z'$ with a mass of $\sim 40~{\rm TeV}$ \cite{fcchh}, we naively expect $m_\rho = {\cal O}(10)~{\rm TeV}$ can be probed by such a future collider but a detailed study is required to have a prospect on that.

\subsection{Gravitational waves}
Since the current model has the confining QCD-like sector, our universe experienced the chiral phase transition and the confinement-deconfinement phase transition in the early stage. If either (or both) of these phase transitions is the first order, the GW can be produced, and it could be detected in a near future observation \cite{Schwaller:2015tja, Antipin:2015xia, Kahara:2012yr, Kang:2021epo, Reichert:2021cvs, Morgante:2022zvc}. Although it is quite tough to study the details of the phase transition because of the strong dynamics, some parameter space of QCD-like theory supports first-order phase transitions. See, e.g., ref.~\cite{Bernhardt:2023hpr} and references therein for recent development and situation. By assuming the electroweak interaction has little impact on the phase transition, the property of the phase transition is determined by $N_c$, $N_f$, and the dark quark mass $m$.

According to a treatment in \cite{Caprini:2015zlo, Caprini:2019egz}, the GW signal from the first order phase transition is parametrized by (dimensionless) latent heat $\alpha$, (dimensionless) inverse time of duration of the phase transition $\tilde\beta$, and the bubble wall velocity $v_w$. The sound wave after bubble collision is considered to be the main source of GW and its peak frequency and amplitude of the GW signal from phase transitions are estimated as \cite{Hindmarsh:2015qta, Caprini:2015zlo}
\begin{align}
  f_{\rm peak} &\sim 1.9 \times 10^1~{\rm Hz} \times \left( \frac{\tilde\beta}{10^5} \right) \left( \frac{1}{v_w} \right) \left( \frac{T}{1~{\rm TeV}} \right) \left( \frac{g_*}{100} \right)^{1/6}, \\
  \Omega_{\rm GW,peak} h^2 &\sim 2.65 \times 10^{-16} \times \left(\frac{\tilde\beta}{10^5}\right)^{-1} \left(\frac{\kappa^2}{10^{-5}} \right) \left( \frac{\alpha}{1+\alpha} \right)^2 \left( \frac{g_*}{100} \right)^{-1/3} \left( \frac{g_{*,{\rm dark}}}{ g_{*,{\rm dark}} + g_{*,{\rm SM}}} \right)^2.
\end{align}
We can see that smaller $\tilde\beta$ is preferred to have a sizable GW signal.

In ref.~\cite{Pasechnik:2023hwv}, the phase transition in QCD-like theories is explored by employing the Polyakov-loop improved linear sigma model as an effective theory and found that $\tilde\beta$ could be as small as ${\cal O}(10^3)$ when the mass of the sigma meson is small otherwise $\tilde\beta \sim 10^4$ -- $10^5$ (see also ref.~\cite{Reichert:2021cvs}). If this is the case, BBO \cite{Crowder:2005nr, Corbin:2005ny, Harry:2006fi} and DECIGO \cite{Seto:2001qf, Kawamura:2006up, Yagi:2011wg, Isoyama:2018rjb} are reachable $\Omega_{\rm GW} h^2 \sim {\cal O}(10^{-17})$ at $f \sim {\cal O}(0.1)$ Hz, and we could have a chance to detect the GW.
This feature of relatively small $\tilde\beta$ is shared with scenarios with a light dilaton \cite{Witten:1980ez, Antipin:2012sm, Hambye:2013dgv, Sannino:2015wka, Ellis:2020awk, Chishtie:2020tze, Baldes:2021aph, Sagunski:2023ynd}.
For given $N_c$, there exists a range of $N_f$ in which the QCD-like theories have a non-trivial IR fixed point. We can expect that a light dilaton appears if $N_f$ becomes close to this conformal window and we could have a detectable GW signal. 
In addition to this, the dark quark mass $m$ gives non-trivial effects on the dynamics of the phase transition (See, e.g., figure~1 in ref.~\cite{Bernhardt:2023hpr}).
To summarize, whether we can detect the GW signals from composite DM models or not is highly non-trivial,
and a dedicated study is required to have a conclusive answer.

\section{Conclusion}\label{sec:conclusion}
We have discussed a model of composite DM based on a new QCD-like sector with $SU(N_c)$ gauge symmetry. For concreteness of numerical analysis, we have focused on the case where $N_c  =3$ and three flavors of dark quarks identified as an $SU(2)_W$ triplet with $Y=0$. The dark sector is confined and its chiral symmetry is spontaneously broken. $G$-parity odd dark pion $\chi$ and even dark pion $\pi$ arise from this symmetry breaking, and the lightest component of $\chi$ and the lightest dark baryon $B$ are candidates of the DM.

The renormalizable Lagrangian allows the mass term of the dark quark and the $\theta$-term in the dark sector,
and we have discussed their impact on DM physics by utilizing the chiral Lagrangian.
In the massless dark quark limit, the electroweak radiative correction induces the mass splitting between $\chi$ and $\pi$ and their mass ratio is fixed to be $m_\pi / m_\chi = \sqrt{3}$. Thus, $\pi$ hardly affects the freeze-out process of $\chi$ and the relic abundance of $\chi$ is determined by their annihilation into the SM gauge bosons as the MDM scenario. On the other hand, nonzero dark quark mass induces the universal mass contribution to both $\chi$ and $\pi$ and their masses tend to be degenerate. We have found, for $m \gtrsim 0.1~{\rm TeV}$, $\pi$ also affects the relic abundance of $\chi$ via the forbidden annihilation channel $\chi\chi\to\pi\pi$ and the mass of $\chi$ to give $\Omega_{\rm DM} h^2 = 0.12$ becomes larger than the mass of the MDM scenario.
Thus, the current model naturally accommodates the forbidden dark matter scenario, and we found our setup realizes the DM mass with ${\cal O}(1$--$100)~{\rm TeV}$.
The $\theta$-term in the dark sector has CP-violating effect, and the current model predicts the electron EDM as $\sim 10^{-32}~e~{\rm cm}$ as a maximal value. This value is much below the current constraints but it can be a good benchmark point for future electron EDM experiments.

The current model has a simple structure but it provides a rich phenomenology as we have discussed in section~\ref{sec:signals}. Although the DM mass is larger than $\sim 1~{\rm TeV}$, The near future progress in indirect detection experiments will increase the chance of observing the annihilation of the DM in our galaxy. Detailed discussion requires careful study of the Sommerfeld effect and we leave it for future study. In addition, we could test this model at the future 100 TeV $pp$ collider, direct detection experiments, and GW observations.
In closing, we comment that our analysis can be easily extended to more generic models. Confining gauge interaction can be, for example, $SO(N_c)$ or $Sp(2N_c)$ gauge interaction. We can also take other SM charge assignments on the dark quarks. In addition to this, a flavor singlet pseudoscalar $\eta'$ can also play a role of $G$-parity even pions in large $N_c$ limit. $\eta'$ can be interpreted as an NG boson from $U(1)_A$ breaking. Although it obtains mass from the chiral anomaly, this contribution is suppressed by $1/N_c$ \cite{Veneziano:1979ec, Witten:1979vv}, and the mass splitting between $\eta'$ and pions is suppressed in large $N_c$ limit.

\section*{Acknowledgements}
The authors thank Kohei Fujikura, Kunio Kaneta, and Satoshi Shirai for useful comments and discussions.
We draw Feynman diagrams using \texttt{TikZ-Feynhand} \cite{Dohse:2018vqo} and \texttt{TikZ-Feynman} \cite{Ellis:2016jkw}.
The work is supported in part by JSPS KAKENHI Grant Numbers 21K03549 (T.A.), 23K03415(R.S.), 24H02236(R.S.), and 24H02244(R.S.).

\appendix

\section{Wess-Zumino-Witten term}\label{sec:wzw}
In this appendix, we discuss the Wess-Zumino-Witten term. For details, see, e.g., section VII-5 of ref.~\cite{Donoghue:2022wrw}. Here we consider a QCD-like $SU(N_f)$ gauge theory with $N_f$ flavors of quarks.
Let us write $U = \exp(2i\varphi)$, and $\ell_\mu$ and $r_\mu$ are external currents which couples to left-handed and right-handed fermions, respectively. $\varphi$, $\ell_\mu$, and $r_\mu$ are $N_f \times N_f$ matrix fields. Then, the Wess-Zumino-Witten term is given as
\begin{align}
{\cal L}_{\rm WZW}
&= - \frac{N_c}{4\pi^2} \epsilon^{\mu\nu\rho\sigma} \int_0^1 d\tau {\rm tr}\Biggl[\varphi \biggl(
\frac{1}{4} \bar v_{\mu\nu} \bar v_{\rho\sigma}
+ \frac{1}{12} \bar a_{\mu\nu} \bar a_{\rho\sigma} \nonumber\\
& \qquad\qquad\qquad\qquad
-\frac{2i}{3} \bar a_\mu \bar a_\nu \bar v_{\rho\sigma} - \frac{2i}{3} \bar a_\mu \bar v_{\nu\rho} \bar a_\sigma -\frac{2i}{3} \bar v_{\mu\nu} \bar a_\rho \bar a_\sigma - \frac{8}{3} \bar a_\mu \bar a_\nu \bar a_\rho \bar a_\sigma
\biggr) \Biggr], \label{eq:generic wzw}
\end{align}
where
\begin{align}
\xi_\tau &\equiv \exp(i\tau\varphi),\\
\bar \ell_\mu &\equiv \xi_\tau^\dagger \ell_\mu \xi_\tau - i \xi_\tau^\dagger \partial_\mu \xi_\tau, \quad
\bar \ell_{\mu\nu} \equiv \xi^\dagger_\tau ( \partial_\mu \ell_\nu - \partial_\mu \ell_\nu + i [\ell_\mu, \ell_\nu] ) \xi_\tau, \\
\bar r_\mu &\equiv \xi_\tau r_\mu \xi_\tau^\dagger - i \xi_\tau \partial_\mu \xi_\tau^\dagger, \quad
\bar r_{\mu\nu} \equiv \xi_\tau ( \partial_\mu r_\nu - \partial_\mu r_\nu + i [r_\mu, r_\nu] ) \xi_\tau^\dagger, \\
\bar v_\mu &\equiv \frac{1}{2}\bar\ell_\mu + \frac{1}{2}\bar r_\mu, \quad
\bar v_{\mu\nu} \equiv \frac{1}{2}\bar\ell_{\mu\nu} + \frac{1}{2}\bar r_{\mu\nu}, \\
\bar a_\mu &\equiv \frac{1}{2}\bar\ell_\mu - \frac{1}{2}\bar r_\mu, \quad
\bar a_{\mu\nu} \equiv \frac{1}{2}\bar\ell_{\mu\nu} - \frac{1}{2}\bar r_{\mu\nu}.
\end{align}

Let us apply this generic expression to the model described in section~\ref{sec:models}.
Then, we identify
\begin{align}
\varphi \to \frac{1}{\sqrt{2} f_d} (\Pi_3 + \Pi_5), \quad
\ell_\mu \to W_\mu, \quad
r_\mu \to W_\mu. \label{eq:wzw fields}
\end{align}
Here $W_\mu$ is the matrix form of the $SU(2)_W$ gauge field defined in eq.~(\ref{eq:Wmu matrix}).
Then, we can rewrite $\bar a_\mu$, $\bar v_{\mu\nu}$, and $\bar a_{\mu\nu}$ by using $W_\mu$ as
\begin{align}
\bar a_\mu &= \frac{1}{2} \xi_\tau^\dagger W_\mu \xi_\tau - \frac{1}{2} \xi_\tau W_\mu \xi_\tau^\dagger - i \xi_\tau^\dagger \partial_\mu \xi_\tau + i \xi_\tau \partial_\mu \xi_\tau^\dagger, \label{eq:amubar}\\
\bar v_{\mu\nu} &= \frac{1}{2} \xi_\tau^\dagger W_{\mu\nu} \xi_\tau + \frac{1}{2} \xi_\tau W_{\mu\nu} \xi_\tau^\dagger, \label{eq:vmunubar}\\
\bar a_{\mu\nu} &= \frac{1}{2} \xi_\tau^\dagger W_{\mu\nu} \xi_\tau - \frac{1}{2} \xi_\tau W_{\mu\nu} \xi_\tau^\dagger. \label{eq:amunubar}
\end{align}
Under the $G$-parity transformation defined in eqs.~\eqref{eq:gparity elementary} and \eqref{eq:gparity}, $\varphi$ and $\xi_\tau$ transform as
\begin{align}
\varphi \to \varphi^T, \quad
\xi_\tau \to \xi_\tau^T.
\end{align}
Since $W_\mu$ is invariant under the $G$-parity transformation, 
we obtain the transformation law of $\bar a_\mu$, $\bar v_{\mu\nu}$, and $\bar a_{\mu\nu}$ given in eqs.~\eqref{eq:amubar}--\eqref{eq:amunubar} as
\begin{align}
\bar a_\mu \to \bar a_\mu^T, \quad
\bar v_{\mu\nu} \to -\bar v_{\mu\nu}^T, \quad
\bar a_{\mu\nu} \to \bar a_{\mu\nu}^T.
\end{align}
Thus, ${\cal L}_{\rm WZW}$ given in eq.~(\ref{eq:generic wzw}) is invariant under this transformation.

At the leading order of the pion fields, we can approximate as $\bar\ell_\mu \simeq \tau \partial_\mu \varphi$ and $\bar r_\mu \simeq -\tau \partial_\mu \varphi$.
Then, we obtain
\begin{align}
\bar a_\mu \simeq \tau \partial_\mu \varphi, \quad
\bar v_{\mu\nu} \simeq W_{\mu\nu}, \quad
\bar a_{\mu\nu} \simeq 0.
\end{align}
By plugging those expressions into eq.~(\ref{eq:generic wzw}), we obtain eq.~(\ref{eq:wzw}).

\bibliography{ref}
\bibliographystyle{JHEP}

\end{document}